\documentclass[twocolumn,oldversion]{aa}
\usepackage{graphicx}
\usepackage{epsfig}
\usepackage{color}
\usepackage{txfonts}
\usepackage{natbib}
\usepackage{longtable}
\bibpunct{(}{)}{;}{a}{}{,} 

\newcommand{\as}[2]{$#1''\,\hspace{-1.7mm}.\hspace{.1mm}#2$}

\newcommand{\Msunperyrperarea}{\mbox{M$_{\sun}$ yr$^{-1}$ kpc$^{-2}$}}

\newcommand{\HI}{\mbox{H{\sc i}}}
\newcommand{\HII}{\mbox{H{\sc ii}}}
\newcommand{\Ha}{\mbox{H$\alpha$} }

\newcommand{\NII}{\mbox{[N{\sc ii}]} }

\newcommand{\SII}{\mbox{[S{\sc ii}]}}
\def\approxlt{\lower.2em\hbox{$\buildrel < \over \sim$}}
\def\approxgt{\lower.2em\hbox{$\buildrel > \over \sim$}}

\def\gtrsim{\mathrel{\hbox{\rlap{\hbox{\lower4pt\hbox{$\sim$}}}\hbox{$>$}}}}
\newcommand{\kms}{\mbox{{\rm km}\,{\rm s}$^{-1}$}}

\def\lesssim{\mathrel{\hbox{\rlap{\hbox{\lower4pt\hbox{$\sim$}}}\hbox{$<$}}}}
\newcommand{\lsim}{\stackrel{<}{_{\sim}}}

\def\la{\mathrel{\hbox{\rlap{\hbox{\lower4pt\hbox{$\sim$}}}\hbox{$<$}}}}
\def\ga{\mathrel{\hbox{\rlap{\hbox{\lower4pt\hbox{$\sim$}}}\hbox{$>$}}}}

\begin{document}

\authorrunning{Lehnert et al.}

\title{On the self-regulation of intense star-formation in galaxies at
z=1-3\thanks{Data obtained as part of program IDs: 074.A-9011, 075.A-0318,
075.A-0466, 076.A-0464, 076.A-0527, 076.B-0259, 077.B-0079, 077.B-0511,
078.A-0055, 078.A-0600, 079.A-0341 and 079.B-0430 at the ESO-VLT.}}

\titlerunning{Self-regulated star formation in z$\sim$2 galaxies}

\author{M. D. Lehnert\inst{1,2}, L. Le Tiran\inst{1},
N. P. H. Nesvadba\inst{3}, W. van Driel\inst{1}, F. Boulanger\inst{3},
\and P. Di Matteo\inst{1}}
\institute{GEPI, Observatoire de Paris, UMR 8111, CNRS, Universit\'e Paris Diderot, 5 place Jules Janssen, 92190 Meudon, France
\and
Institut d'Astrophysique de Paris, UMR 7095, CNRS, Universit\'e Pierre et Marie Curie, 98 bis Bd Arago, 75014 Paris, France
\and 
Institut d'Astrophysique Spatiale, UMR 8617, CNRS, Universit\'e Paris-Sud, B\^atiment 121, 91405 Orsay Cedex, France}

\date{Accepted, Received}

\abstract{We have analyzed the properties of the H$\alpha$ and [N{\sc
ii}]$\lambda$6583 rest-frame optical emission lines of a sample of 53
intensely star forming galaxies at z=1.3 to 2.7 observed with SINFONI
on the ESO-VLT. Similar to previous authors, we find large velocity
dispersions in the lines, $\sigma$=few 10-250 km s$^{-1}$. Our data
agree well with simulations where we applied beam-smearing and assumed
a scaling relation of the form: velocity dispersion is proportional to
the square root of the star-formation intensity (star-formation rate
per unit surface area). We conclude that the dispersions are primarily
driven by star formation. To explain the high surface brightness and
optical line ratios, high thermal pressures in the warm ionized medium,
WIM, are required (P/k$\ga$10$^6$--10$^7$ K cm$^{-3}$). Such thermal
pressures in the WIM are similar to those observed in nearby starburst
galaxies, but occur over much larger physical scales. Moreover, the
relatively low ionization parameters necessary to fit the high surface
brightnesses and optical line ratios suggest that the gas is not only
directly associated with regions of star formation, but is wide spread
throughout the general interstellar medium. Thus the optical emission
line gas is a tracer of the large scale dynamics of the bulk of the ISM.

We present a simple model for the energy input from young stars in an
accreting galaxy, to argue that the intense star-formation is supporting
high turbulent pressure, which roughly balances the gravitational pressure
and thus enables distant gas accreting disks to maintain a Toomre disk
instability parameter Q$\sim$1. For a star formation efficiency of  3\%,
only 5-15\%  of the mechanical energy from young stars that is deposited
in the ISM is needed to support the level of turbulence required for
maintaining this balance. Since this balance is maintained by energy
injected into the ISM by the young stars themselves, this suggests that
star formation in high redshift galaxies is self-regulating.}


\maketitle

\section{Introduction}\label{sec:intro}

Galaxies exhibit a wide range of phenomena, some of which appear to
have been more important in the distant Universe than today. The
global co-moving star-formation rate was higher by an order of
magnitude \citep{madau96}, and so were the specific star-formation
rates \citep{elbaz07, daddi07}. Morphologies were increasingly
irregular, an observation that cannot be explained with extinction
alone \citep{conselice08}. Processes like starburst-driven outflows
are obvious and ubiquitous in samples of high redshift galaxies
\citep[e.g.,][]{shapley03, steidel10} compared to only a small fraction
of galaxies today \citep{lehnert96}. The merger rate is also likely to be
much higher \citep{ravel09}. The gas infall rates of nearby galaxies are
apparently low, but they were likely much higher in the early Universe
\citep{croton06}. Today, the Universe can be considered in the ``age
of secular evolution'', but until about 8 Gyrs ago it was in the ``age
of feedback and self-regulation''. It is the mix of physical processes
during their early evolution that shaped the characteristics of the
ensemble of the galaxies we observe today.

Understanding this complexity and the relative contribution of the
physical processes that shaped distant galaxies is required before we can
consider our understanding of galaxies in any sense complete. With the
higher activity levels of galaxies at z$\sim$1-5 relative to today it is
difficult to gauge the importance of each process by mere analogy. This
requires careful, direct observations of distant galaxies.

Of significant concern in this pursuit is that obtaining detailed spatial
information from distant galaxies is hampered by the large impact of
cosmological surface brightness dimming, e.g. by a factor $\sim$150 for
z=2.5. This implies that only the highest surface brightness galaxies
are spatially resolved with integral field spectroscopy or in continuum
imaging. It also biases the samples where we can detect spatially extended
emission line gas to galaxies which have intense star formation over many
tens of kpc$^2$ or extended emission line regions excited by powerful
active galactic nuclei \citep[e.g.,][]{nesvadba06, nesvadba08}.

Observations of intensely star-forming galaxies at z$\sim$2 with bright,
extended line emission reveal remarkably broad line widths implying
important random, and presumably (at least partially) turbulent,
gas motions. Limited by the $\sim$ 1-few kpc resolution of current
spectrographs, the observed line widths reflect the blended bulk
and turbulent gas motions on kpc scales. In galaxies where velocity
gradients are spatially resolved, both motions have very similar
amplitudes, unlike in nearby galaxies where line widths are much
smaller than circular velocities \citep{law07,law09,fs09, epinat09,
epinat10, epinat12}. This requires either a more intense, or perhaps
more efficient, source or an additional source of turbulent energy at
high redshift. The possible sources for  energizing the turbulence
remain the subject of significant controversy. Some authors suggest
that the turbulent cascade is primarily initiated by shear and large
scale gravitational or fluid instabilities, perhaps further energized by
cosmological gas accretion \citep{Brooks09,Keres09,Dekel09,elmegreen10},
while others give a significant role to the energy injection from young
stars \citep[e.g.,][]{elmegreen04, cox05, ferrara93, norman96}. However,
the dichotomy between gravitational and stellar sources of energy is
largely artificial as both no doubt play important roles. Supersonic
turbulence can provide global support  to the ISM, but it naturally
produces density enhancements that may collapse locally and may
subsequently form stars or dissolve. Global star formation in galaxies
appears to be controlled by the balance between gravity and turbulence
\citep{maclow04}. This balance is moderated by the importance of each
form of energy injection, which on the largest scales depends on the
rate and scale at which it is injected, and on how rapidly the energy
cascades and dissipates in various phases, and over what time and size
scales \citep[e.g.,][]{elmegreen04}. None of these variables are well
constrained in distant galaxies. To some extent, the global regulation
of star formation mirrors that taking place on the smallest scales,
where the dissipation of turbulent energy in self-gravitating clouds is
important for regulating star formation within these clouds.

\citet{L09} found that the velocity dispersions, $\sigma$, of optical
emission-line gas in intensely star-forming galaxies at z$\sim$2 scale
with the star-formation rate per unit area (star-formation intensity,
$\Sigma_{\rm SFR}$) as $\sigma$ $\propto$ $\Sigma_{\rm SFR}^{0.5}$
\citep[see also][]{swinbank12, MD13}. They suggested that at a very
simple level, such a relationship would be expected if the lines
were broadened by the mechanical energy injection from young stars
\citep{dib06}. Subsequently, \citet{Green10} came to similar conclusions
from observing strongly star-forming galaxies at low redshifts which
had similar overall H$\alpha$ luminosities as galaxies observed at
z$\sim$2 with integral field spectroscopy \citep{L09}. More recently,
and through an analytical approach, \citet{ostriker11} found that such a
relationship might be the natural outcome of dense ($\Sigma_{\rm gas}$
$\sim$few 100 M$_{\sun}$ pc$^{-2}$), star-forming gas disks which are
regulated through the momentum input from star formation. Compared to the
other possible sources such as gas accretion, star-formation intensity in
high-z galaxies can be constrained observationally, e.g. by measuring
H$\alpha$ luminosities and spatial distributions, although heavily
extincted regions may have their intensities underestimated.

A number of authors subsequently pointed out that this relationship
is also consistent with the Toomre disk stability criterion
\citep[e.g.,][]{krumholz10}. Although not discussed in their work,
a Toomre parameter Q$\sim$1 characteristic of gas disks which are
marginally unstable against star formation leads to a relationship of
the type observed, $\sigma$ $\propto$ $\Sigma_{\rm SFR}^{0.6-0.7}$, if
these galaxies follow the Schmidt-Kennicutt relation. \citet{burkert10}
develop this idea further by arguing that disks must stay close to the
gravitational stability line. They suggest that large scale gravitational
instabilities will generate density and velocity irregularities which
drive turbulence and heat the gaseous disk. This process saturates
close to the instability line and thus, \citet{burkert10} argue that
gravitational forcing is sufficient to explain the highly turbulent
gas. \citet{elmegreen10} find that their analytical model of intensely
star-forming, fragmenting disks can reproduce the observed velocity
dispersions in high redshift galaxies if Q is set equal to  1, suggesting
that whatever the mechanism responsible for driving galaxies towards
the line of instability, it can reproduce the observations.

The basic difficulty is that observing velocity gradients and line
widths consistent with marginally Toomre-stable disks only implies
that distant star-forming galaxies are in a state where they can form
stars intensely, but it does not purport to explain the source of the
turbulence. Galaxies at z$\sim$2 appear to maintain their high gas
turbulence over long star-formation time scales of several $10^8$ yrs,
and maintaining Q$\sim$1 over similar timescales is only possible if
turbulent dissipation is balanced by the injection of turbulent energy
over similar length of time. Since turbulent energy dissipates rapidly,
on only a dynamical time \citep[][]{maclow99}, this seems difficult
to ensure with gravity alone over the extended star-formation periods
\citep{elmegreen10}. Moreover, gravitationally-driven processes alone
are perhaps insufficient for producing even the relatively modest
dispersion observed in the warm neutral and ionized medium of nearby
galaxies \citep{tamburro09, leroy08}. Correlations between star-formation
rate and \HI\ velocity dispersions in nearby galaxies \citep{tamburro09}
as well as the decrease in CO line dispersions with radius in some nearby
galaxies \citep{wilson11} favor an important, but perhaps not dominant,
contribution from the energy injection by  supernovae and stellar
winds in regulating the turbulence of the ISM in nearby galaxies. The
same also appears true for intensely star-forming relatively nearby
galaxies with properties similar to high redshift Lyman break galaxies
 \citep{basu-zych09, goncalves10}.

Here we present new observational data on how star formation intensities
and gas kinematics in high-redshift galaxies are related, extending our
initial sample from 11 galaxies to 53, and covering a larger redshift
range. Apart from putting our previous claims on statistically more
robust grounds this also allows us to demonstrate the effects of
surface-brightness dimming directly. We discuss the inevitable biases
in all studies of star formation in high-redshift galaxies related
to surface-brightness dimming and beam-smearing from our data, and
demonstrate that our results are not dominated by these effects. With
this larger sample we confirm the results of \citet{L09} that the ISM in
the observed high-redshift galaxies has very high pressures compared to
galaxies at low redshift. We argue that this enables star formation in
these galaxies to be self-regulated, and demonstrate that self-regulation
leads very naturally to a Toomre parameter, Q$\sim$1, generalizing the
analytical model first presented by \citet{elmegreen10}. We further
paint a picture of what the optical emission line gas may be showing
us about the turbulent cascade and its impact on star formation in high
redshift galaxies.

Throughout the paper we adopt a flat H$_0 =$70 km s$^{-1}$ Mpc$^{-3}$
cosmology with $\Omega_{\Lambda} = 0.7$ and $\Omega_{M}= 0.3$.

\section{Source characteristics}\label{sec:observations}

\begin{figure}
\includegraphics[width=9.0cm]{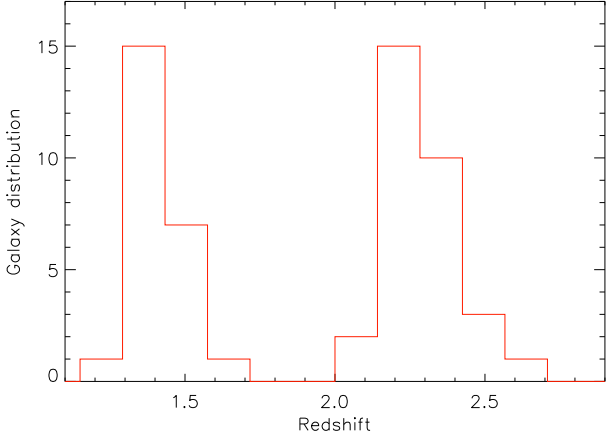}
\caption{Redshift distribution of our sample of galaxies. The two
observed redshift ranges of the H$\alpha$ and \NII$\lambda$6583 lines
(z$<$1.8 and z$>$2)are due to atmospheric transmission and the limited
wavelength coverage of the H- and K-bands.}
\label{fig:histo_z}
\end{figure}

Our sample of 53 galaxies was observed with SINFONI on the
ESO-VLT.  Their basic properties of the 53 galaxies are presented in
Table~\ref{table:properties}. They span a redshift range of z=1.3-2.7
(Fig~\ref{fig:histo_z}). For details of the observations of the galaxies
in this sample see the summaries in e.g. \citet{L09, fs06, fs09}. All were
reduced by us in a similar way as previously described by \citet[][]{L09}.

The data, which were retrieved from the ESO archives, were taken for
a variety of programs and the selection of the galaxies is rather
inhomogeneous. As we are not performing a population study, for the
purposes of this paper, this is adequate.  The key to our sample is that
all galaxies have sufficiently high H$\alpha$ surface brightness to obtain
spatially resolved line maps in a few hours integration time (which is
typical for SINFONI observations of distant galaxies). Because of this,
our results are only applicable to galaxies with high star-formation
intensity (star-formation rate per unit area) which necessitates a
process that is able to drive intense star-formation for 100 Myrs or
more \citep{erb06} and my introduce a bias. However, it should
be pointed out that the characteristics of the ensemble of galaxies
studied here are similar to those of the general distant star-forming
galaxy population.  Specifically, the galaxies in our sample for which
the specific star formation rate has been estimated \citep[a majority;
see ][and references therein]{fs09,fs11,genzel11,vergani12}, lie along
or near the ``main sequence of star formation'' for redshifts between
1 and 2 \citep[e.g.][]{elbaz07, daddi07, elbaz11}.

The observations reach surface brightness detection limits in H$\alpha$
of $\sim$2$\times$10$^{-19}$ erg s$^{-1}$ cm$^{-2}$ pixel$^{-1}$ (for
the 125~mas pixel$^{-1}$ scale and averaged over 3$\times$3 pixels;
``mas'' is milliarcseconds). The spectral resolution is FWHM$\sim$115 and
$\sim$150 km s$^{-1}$ in the K and H bands, respectively. The observed
surface brightnesses range from about 3.6-34 $\times$ 10$^{-18}$ erg
cm$^{-2}$ s$^{-1}$ arcsec$^{-2}$ and the objects have isophotal radii,
at their surface brightness detection limits, of 1-2 arcsec$^2$,
or on average $\sim$7$\pm$2 kpc (corresponding to an isophotal area
of $\sim$150$\pm$40 kpc$^2$). The point-spread function FWHM of the
data is $\sim$\as{0}{6}, which at z=2 represents an area of $\sim$20
kpc$^2$, so we have generally $\sim$8 spatial resolution elements per
object. With such low spatial resolution, it is necessary to discuss the
impact of beam smearing on the characteristics of our data (discussed
in detail in \S~\ref{sec:BSanalysis}). In the present analysis, we will
formally define the isophotal radius, r$_{\rm iso}$, as r$_{\rm iso}$ =
(A$_{\rm iso}$/$\pi$)$^{1/2}$, where A$_{\rm iso}$ is the isophotal area,
i.e. the total area of all pixels above the 3-$\sigma$ surface brightness
limit of the data (Table~\ref{table:properties}).

The total H$\alpha$ luminosities of all galaxies in our sample are
above 10$^{41}$ erg s$^{-1}$ and most are above $\sim$10$^{42}$
erg s$^{-1}$. H$\alpha$ luminosities in this range imply an average
star-formation rate of $\sim$70 M$_{\sun}$ yr$^{-1}$. However, as these
are not corrected for extinction our estimates are lower limits. For a
sub-sample of these galaxies we have H$\beta$ estimates, and the line
ratios of H$\alpha$ to H$\beta$ suggest relatively modest extinction
corrections to the star-formation rates, of a factor of few \citep{L09}.

\section{Surface brightness limits and selection effects}\label{sec:SBlimits}

Due to cosmological surface brightness dimming, over the redshift range
of a factor $\sim$2 in our sample there is about a factor of 5 difference
in the faintest rest-frame line emission surface brightness levels that
can be probed, whereas the physical dimensions in the rest frame, i.e.,
length per arcsec, changes by only about 5\%. Given this strong impact
of cosmological surface brightness dimming, it is worth repeating the
conclusion from \cite{L09} that even the lowest redshift sources in our
SINFONI sample are extreme compared to galaxies in the local volume,
where only the most intense starbursts have such high H$\alpha$ surface
brightness, and this on considerably smaller physical scales only.

\begin{figure}
\includegraphics[width=9.0cm]{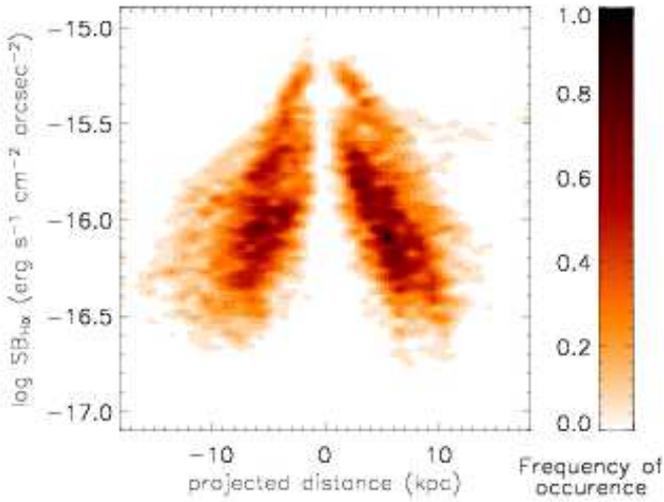}
\caption{A 2-dimensional histogram of the frequency of occurrence of pixel values of H$\alpha$ surface brightness versus projected physical distance for the entire sample. As shown in the color bar, the darkest regions have the highest frequency of occurrence. The surface brightness has not been corrected for cosmological dimming. The zero of the projected distance was chosen to be the symmetry point in either the H$\alpha$ velocity field or of 3-$\sigma$ isophotes in the H$\alpha$ surface brightness map.}
\label{fig:SBvsPos}
\end{figure}

The total range in surface brightness probed by the SINFONI observations
of our ensemble of galaxies is $\approx$20 (Fig.~\ref{fig:SBvsPos}). For
purposes of this comparison and others, we constructed 2-dimensional
histograms which show the frequency of occurrence of various pairs of
values in all pixels of our data sets, after smearing to the spatial
and spectral resolution of the real SINFONI observations -- for example,
surface brightness versus projected distance or \Ha line width versus
star-formation rate intensity. Using the frequency of occurrence is a
clearer way of visualizing various relationships than showing all single
pixels and emphasizes the continuity of the data.

\begin{figure*}
\centering
\includegraphics[width=17.0cm]{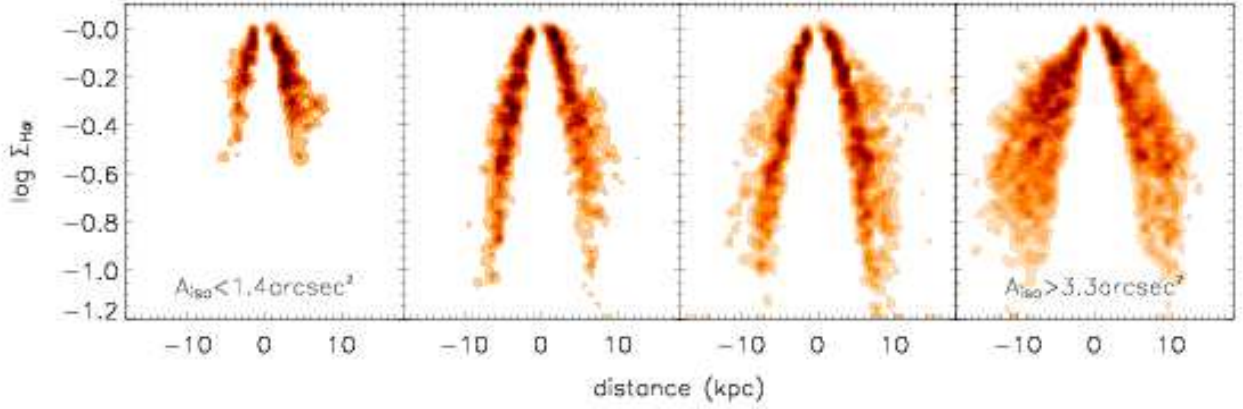}
\caption{Two-dimensional histograms of the relative frequency of
occurrence of normalized H$\alpha$ surface brightness values versus the
projected distance for galaxies in 4 equal-size bins in galaxy number
selected on their isophotal area (A$_{\rm iso}$). The relative frequency
of occurrence is scaled in the same way as in Fig.~\ref{fig:SBvsPos}.
From left to right, each panel represents an increasing angular size
(i.e., A$_{\rm iso}$ $<$ 1.4 arcsec$^2$, 1.4 arcsec$^2$ $\geq$A$_{\rm
iso}$ $<$ 2.5 arcsec$^2$, 2.5 arcsec$^2$ $\geq$A$_{\rm iso}$ $<$
3.3 arcsec$^2$ and A$_{\rm iso}$ $\geq$3.3 arcsec$^2$). The values
of r$_{\rm iso}$, which is derived from A$_{\rm iso}$, for individual
galaxies are given in Table~\ref{table:properties}. The galaxies with
the smaller dynamic range (higher surface brightness detection limits)
in the data are also smaller in isophotal size.}
\label{fig:SBvsPos_Npix}
\end{figure*}

\begin{figure}
\includegraphics[width=9.0cm]{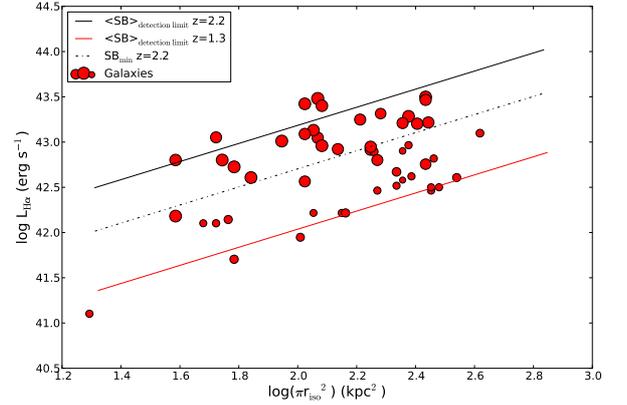}
\caption{Plot of H$\alpha$ luminosities versus isophotal area for the galaxies in our sample (red circles). The size of each circle is proportional to the redshift of the galaxy, with larger circles denoting higher redshifts (range in z is 1.3--2.7). The two solid lines show the relationship between luminosity and area for the average surface brightness detection level of galaxies, 1.1$\times$10$^{-17}$ erg s$^{-1}$ cm$^{-2}$ arcsec$^{-2}$ (black for galaxies at z=2.2 and red for z=1.3, respectively). The dot-dashed line shows the relationship for the minimum surface brightness detection limit of 3.6$\times$10$^{-18}$ erg s$^{-1}$ cm$^{-2}$ arcsec$^{-2}$ for all galaxies with z$\approx$2.2. See
Table~\ref{table:properties} for the \Ha luminosities, isophotal radii and surface brightness detection limits.}
\label{fig:surflims}
\end{figure}

To investigate the dependence of the range of surface brightnesses probed
by our data and its influence on the observed galaxy isophotal size, we
divided our galaxies into 4 bins of equal galaxy number based on their
isophotal areas (Fig.~\ref{fig:SBvsPos_Npix}). For the 3 bins with the
largest average isophotal sizes ($\geq$1.4 arcsec$^2$), the observed
dynamic range in surface brightness goes up to $\sim$20 and is typically
about 15. However, for the apparently smaller galaxies, the dynamic range
of the observations is only about a factor of 3. The relationship between
surface brightnesses and projected distance is the same for all bins,
indicating that their individual declines in surface brightness follow
the same trend. The observations imply two things. First, although the
effects of beam smearing cannot be ignored, as we will discuss extensively
in the next sections, it does not dominate the overall distribution of
the surface brightness (i.e., the distribution does not simply represent
the point-spread-function, PSF, of the observations). Second, it implies
that the reason the apparently small galaxies appear small is that the
surface brightness detection limits of the observations are too high to
detect their extended emission. As noted above, this is not simply due to
the physical scale of the galaxies changing across our redshift range,
as this effect amounts to only $\sim$5\%. While we do not show it here,
there is a crude trend for the galaxies with the highest total \Ha fluxes
to also have the largest isophotal sizes. Again, this suggests that we
are limited by the overall dynamic range of the data and by the impact
of surface brightness dimming, but not by differences in the intrinsic
isophotal sizes of the galaxies (Fig.~\ref{fig:surflims}). This is an
important point to bear in mind when reading this paper.

Furthermore, if the trends in surface brightness and total \Ha luminosity
were dominated by cosmological surface brightness dimming, which is
proportional to (1+z)$^4$, we would expect our data to show a slope
of about 4 in a plot of the logarithmic relation between H$\alpha$
luminosity and isophotal area and indeed, we find such a trend
(Fig.~\ref{fig:surflims}). The range in surface brightness detection
limits in our data trace out the ranges in size and luminosity we observe
in our sample of galaxies. In addition, this confirms that at the high
redshift end of our sample the intrinsic surface brightnesses are very
high and that they generally decline with redshift.

\section{The effect of beam smearing}\label{sec:BSanalysis}

In \citet{L09}, we proposed that the H$\alpha$ emission line velocity
dispersions of the high surface brightness spatially-resolved galaxies
observed with SINFONI (and other near-infrared integral field units)
could be explained by the mechanical output of young stellar populations
within those galaxies. In fact, we framed the argument in the context
of an overly simplified energy injection model of the form, $\sigma$=
($\epsilon$$\Sigma_{\rm SFR})^{1/2}$, where $\sigma$ is the H$\alpha$
velocity dispersion, $\epsilon$ is the coupling efficiency of the
mechanical energy output to the interstellar medium and $\Sigma_{\rm SFR}$
is the star-formation intensity. The energy due to the young stars is
given by population synthesis models \citep{leitherer99}. Using scaling
relations based on models and observations of nearby galaxies, we were
able to explain the trends between dispersion and star-formation intensity
with no free parameters \citep[see ][ for details of the scalings used
in this analysis]{L09}.

\begin{figure*}
\setlength{\unitlength}{6cm}
\begin{picture}(3,1)
\put(0, 0){\includegraphics{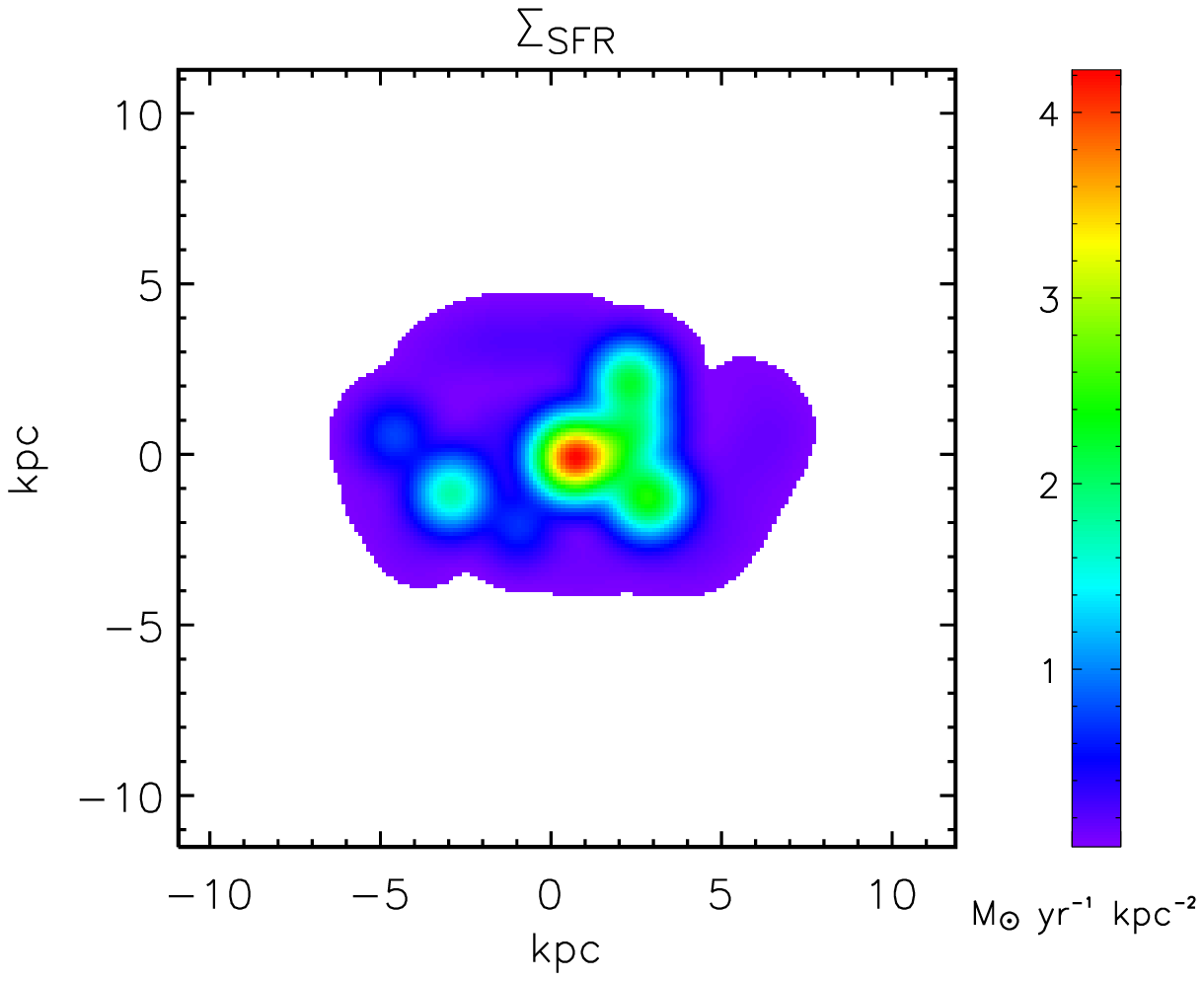}}
\put(1, 0){\includegraphics{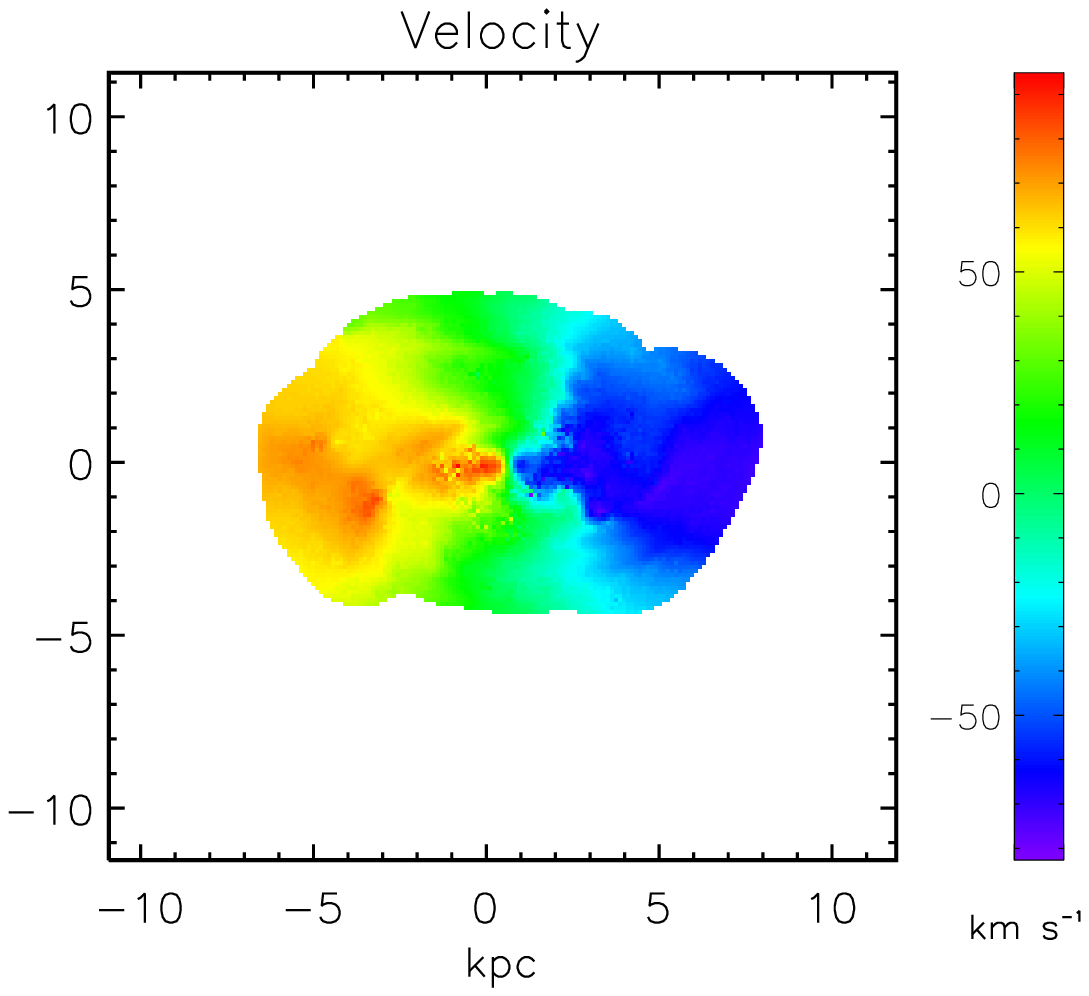}}
\put(2, 0){\includegraphics{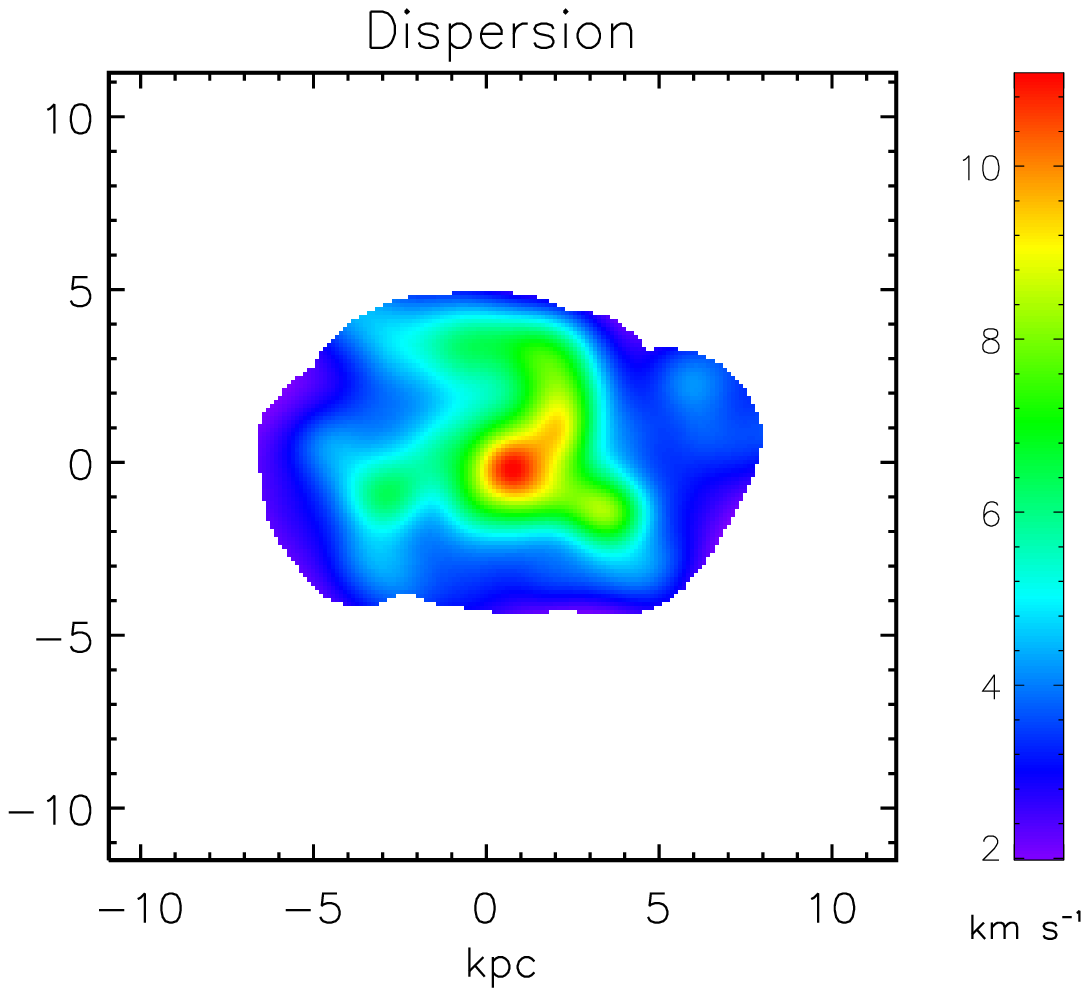}}
\end{picture}
\caption{Output from the simulation of a star-forming disk galaxy
as it would appear at z=2 before it was beam smeared. Shown are the
distribution of the star-formation rate intensity (left panel), \Ha
velocity field (center panel) and \Ha dispersion map (right panel). The
galaxy is projected at an inclination angle of 45$^{\circ}$ relative
to face-on. The scale on the axes is in kpc and the kinematic axis is
horizontal. Despite the different units (kpc vs. arc seconds), we note
that the sizes of these panels and those in Fig.~\ref{fig:analogs}
are similar, for ease of direct comparison.}
\label{fig:simus}
\end{figure*}

However, the limited spatial resolution of integral field spectroscopy
of z$\sim$2 galaxies causes an artificial broadening of the emission
lines due to large-scale kinematics, which can result in a significant
radial velocity gradient across each pixel. While we argued that this
would not have an important impact on our results in \citet{L09} based
on some simple modeling, we now feel it is important to revisit this
issue in a more complete and detailed way.

\subsection{Construction of artificial IFU observations}

In order to study the effect of beam smearing, we used the simulation
of an isolated galaxy already presented in \citet{diMatteo08} which was
made using a Tree-SPH code \citep [see ][ for a description]{seme02,
dimatteo09, chili10}. This model has a high gas mass fraction (50\%)
and the galaxy was evolved in isolation, without companions or tidal
interactions \citep{qu11}. We allowed the disk to go unstable against
star-formation and it developed a ``clumpy'' morphology which evolved
with time \citep[see][for details of the initial conditions and evolution
of the simulation]{diMatteo08}. Intrinsically, our simulations show
low gas velocity dispersions, of-order 10 km s$^{-1}$ (or less), which
are roughly constant with radius and a rotation speed of $\sim$220 km
s$^{-1}$ which implies a v/$\sigma$$\approx$20. From this simulation,
we produced maps of the star-formation intensity, radial velocity field
and velocity dispersion (Fig.~\ref{fig:simus}). For our primary analysis,
the simulation was viewed at an inclination angle of 45$^{\circ}$ and
scaled to have a projected rotational amplitude of 110 km s$^{-1}$,
a value derived from the average intrinsic full width at half maximum
of the integrated spectra of our sample. This implies a rotation speed
of about 160 km s$^{-1}$.

We then degraded the resolution of the simulations to produce synthetic
observations similar to what we observed with SINFONI. We adopted a
single redshift (z=2) for determining the scalings in the synthetic
observations -- the exact redshift chosen does not matter much as the
physical scale per angular projected size (kpc arcsec$^{-1}$) changes
little ($\sim$5\%) over the redshift range of our sources. We also took
into account the instrumental spectral resolution, the seeing using a
FWHM for a Gaussian distributed PSF of \as{0}{6} and added an amount
of noise to the synthetic data consistent with the observations. The
artificial datacubes created in this way were then analyzed using the
same procedures as the real SINFONI observations.

\begin{figure*}
\setlength{\unitlength}{6cm}
\begin{picture}(3,2)
\put(0, 1){\includegraphics{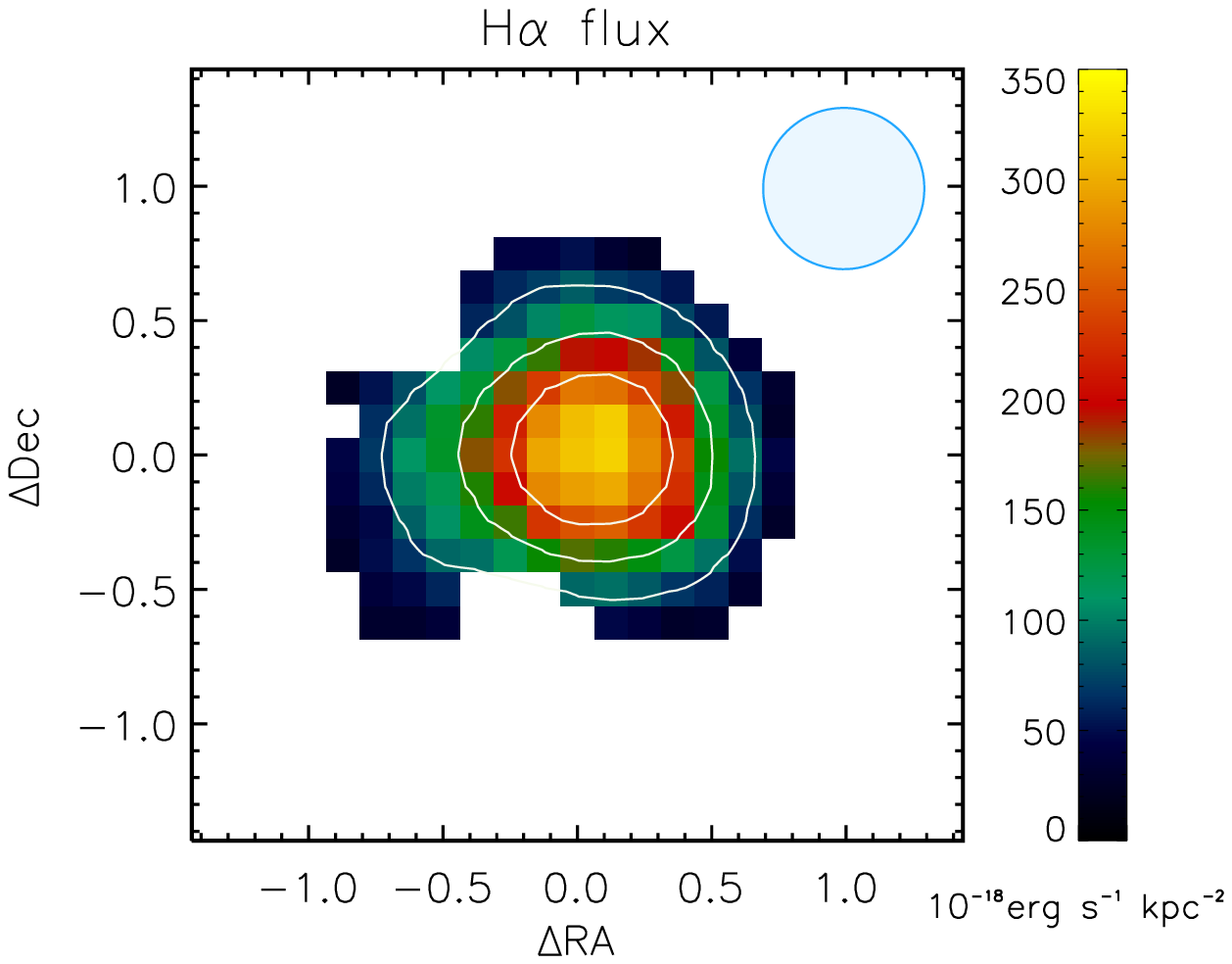}}
\put(1, 1){\includegraphics{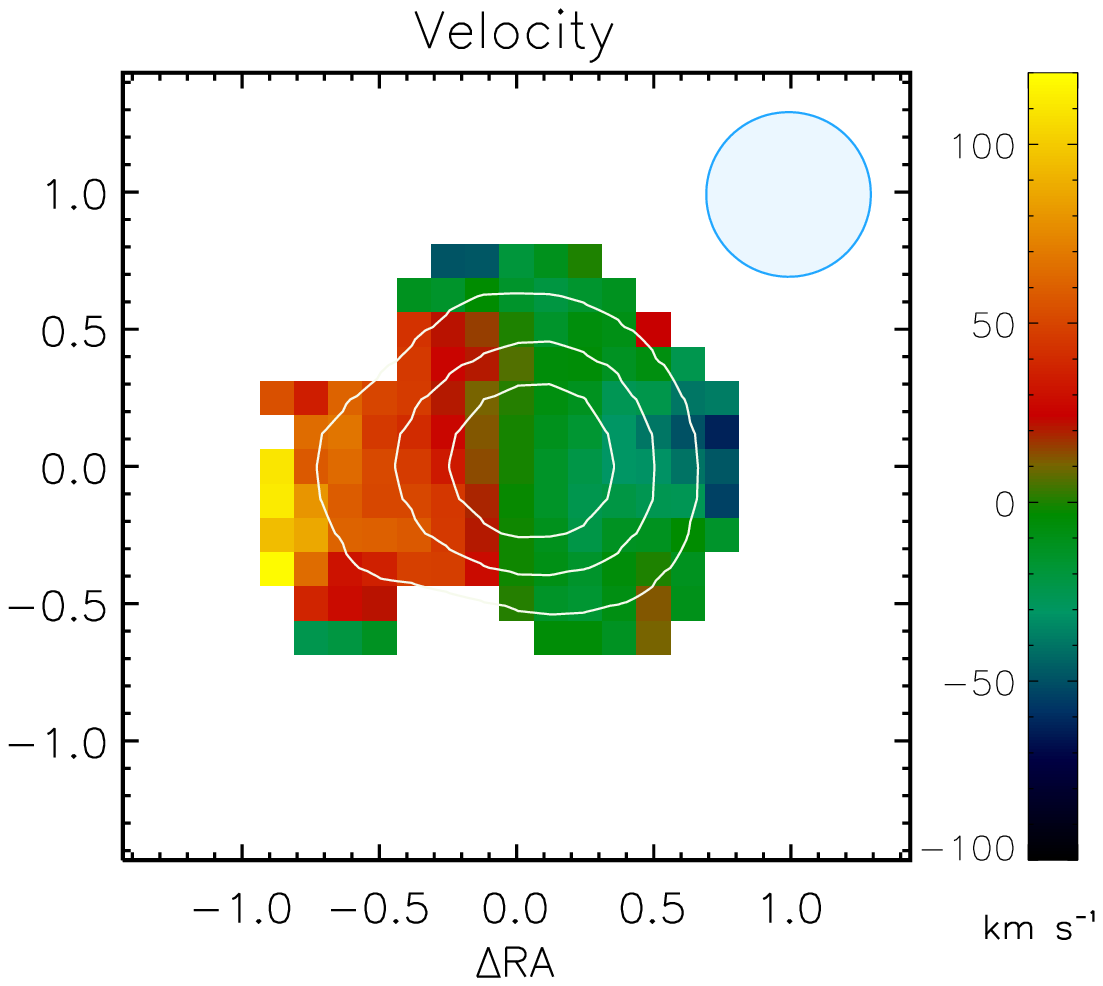}}
\put(2, 1){\includegraphics{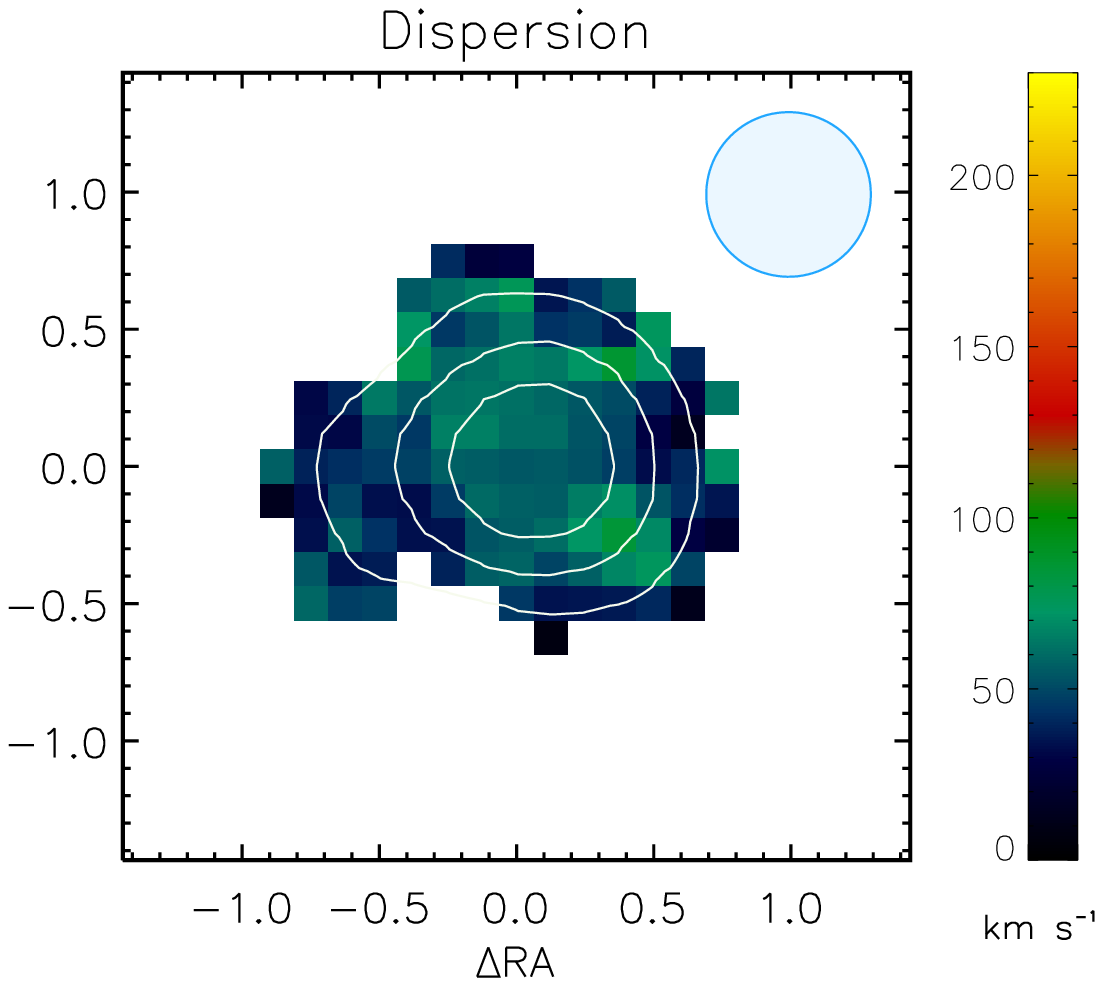}}
\put(0, 0){\includegraphics{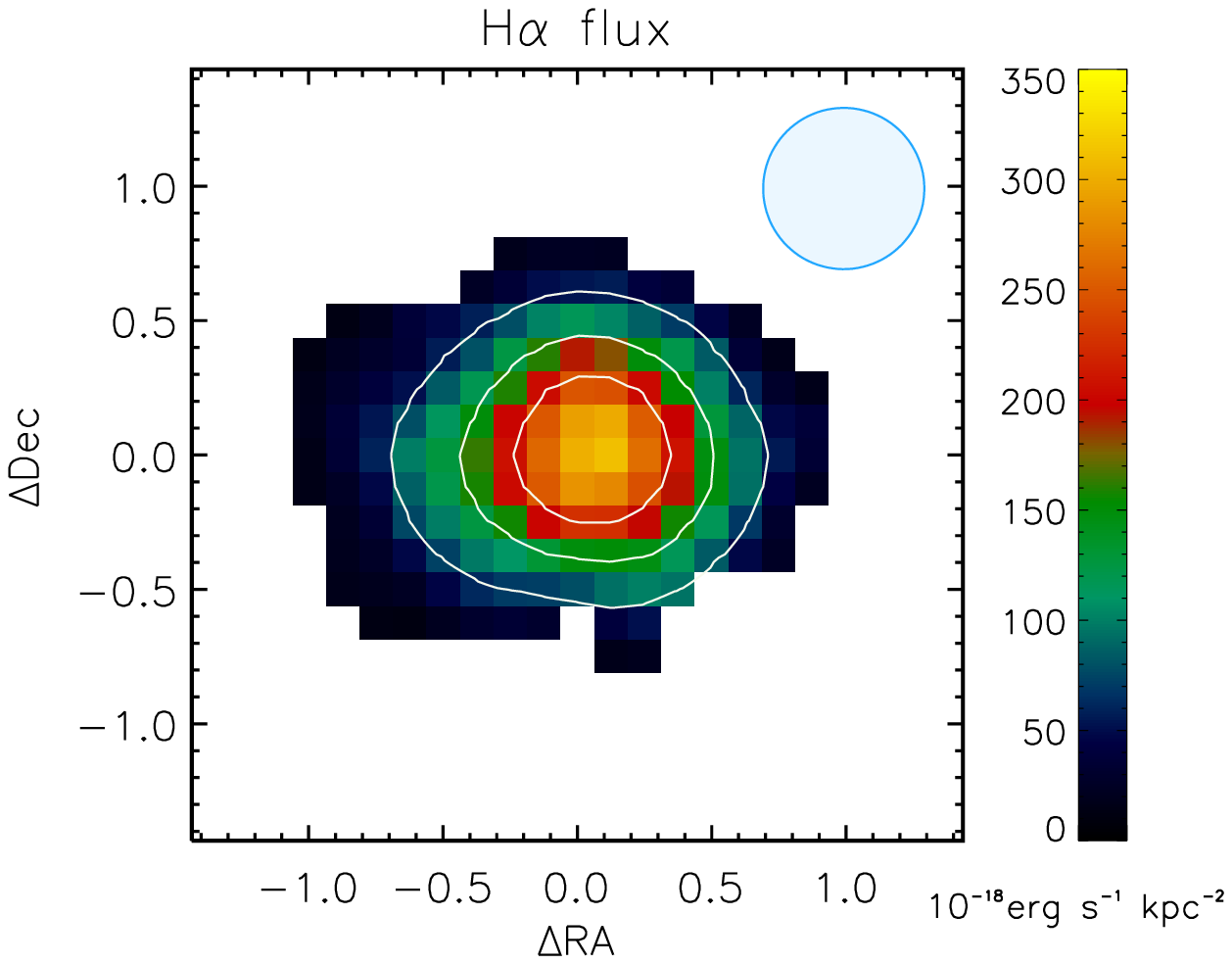}}
\put(1, 0){\includegraphics{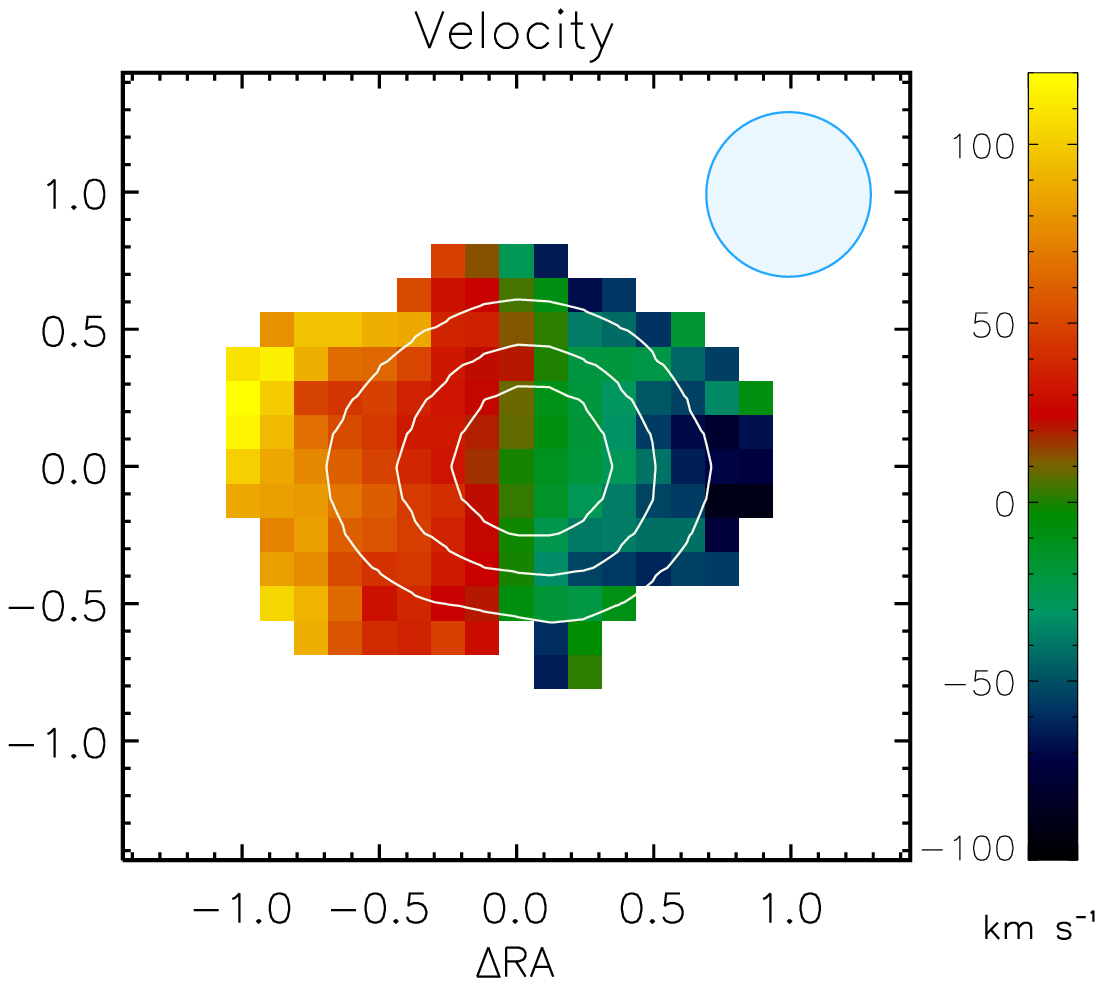}}
\put(2, 0){\includegraphics{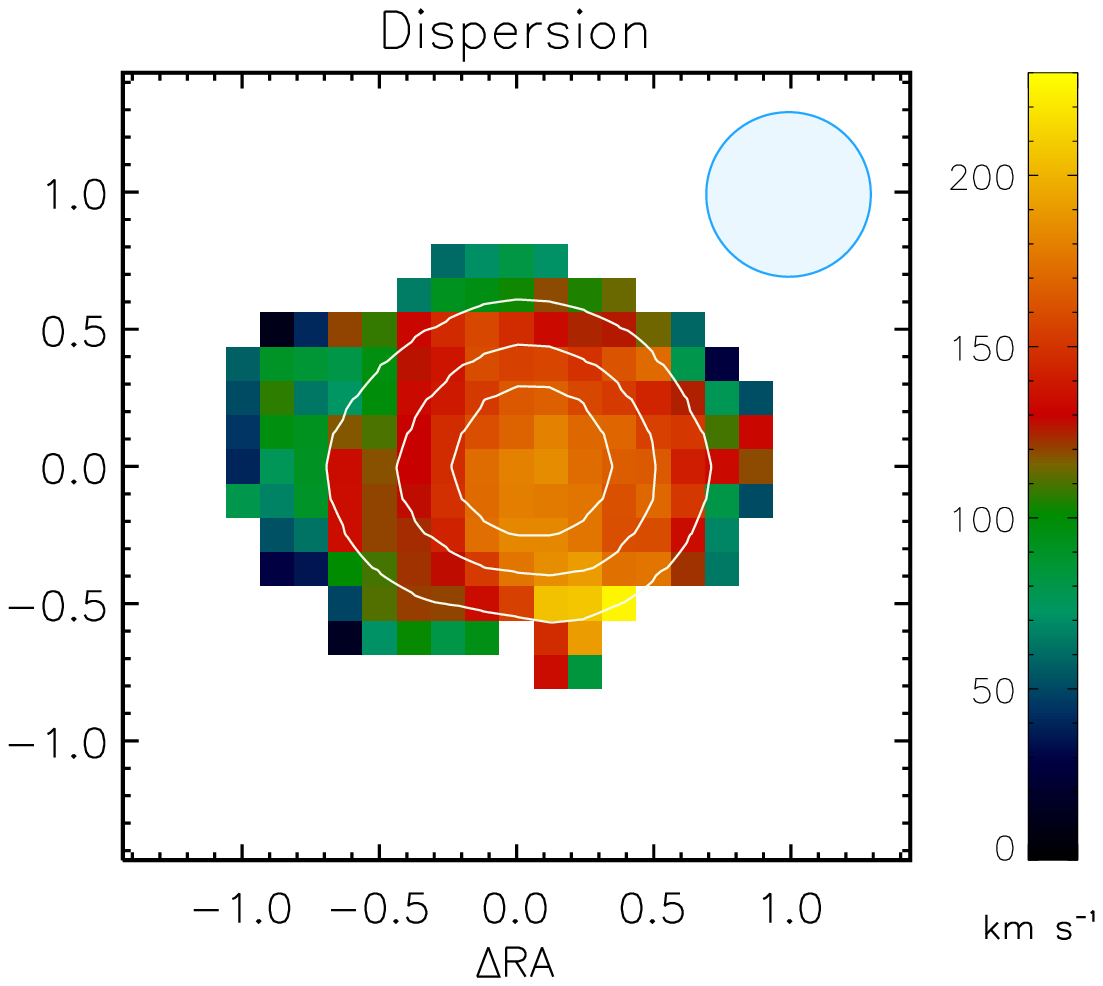}}
\end{picture}
\caption{Results of the simulated data sets of a star-forming disk galaxy
at z=2 after beam smearing using different relationships for the velocity
dispersions. Shown are the distribution of the \Ha surface brightness
(left), velocity field (middle) and velocity dispersion map (right). In
the upper row, we have simply taken the velocity dispersions as estimated
from the N-body/SPH simulation, while for the lower row we assumed a
relationship between the velocity dispersion and star-formation intensity
of the form $\sigma=140\sqrt{}\Sigma_{\rm SFR}$ km s$^{-1}$. The circle
in the upper right of each plot represents the FWHM of the PSF used to
make the artificial data set. The scales are in arc seconds. We note
that we have scaled the images such that the sizes of these panels and
those in Fig.~\ref{fig:simus} are similar, for ease of direct comparison.}
\label{fig:analogs}
\end{figure*}

In the following analysis, we will focus on the impact of beam smearing on
the interpretation of the underlying cause of the high dispersion observed
in distant galaxies with spatially-resolved line emission observations.

In \citealt{L09} we proposed that the dispersion is proportional to
the star-formation intensity. For our analysis of the effects of beam
smearing, we make a comparison between the analogs produced using two sets
of scaling relations for the velocity dispersions: (1) the dispersions
taken directly from the simulation, $\sigma_{\rm sim}$, and (2) a function
of the star formation rate intensity $\sigma=140\sqrt{}\Sigma_{\rm SFR}$
km s$^{-1}$-- see \cite{L09} for details. The resulting \Ha flux and
kinematics maps made with our artificial SINFONI data sets and applying
these two different types of dispersion show, as expected, that the
surface brightness distribution is now smoother than in the original
simulation (Fig.~\ref{fig:analogs}). The star-formation intensities --
the star-formation rate per unit area -- are related to the \Ha surface
brightness by a simple scaling \citep[e.g.][]{kennicutt98}. We use this
scaling to produce the \Ha surface brightness distribution which is
subsequently beam smeared.

\subsection{Effect of beam smearing on star-formation intensities}

\begin{figure}
\includegraphics[width=9.0cm]{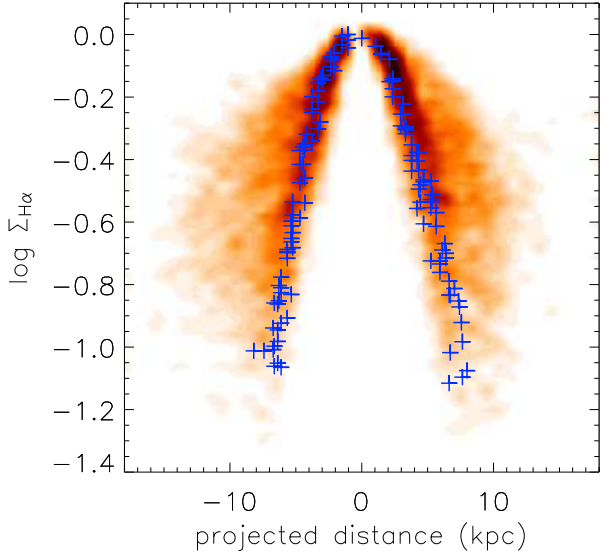}
\caption{Comparison of the relative frequency of occurrence of pixel
values of normalized per pixel \Ha surface brightness distributions of
the whole sample of galaxies (in shades of red) with superimposed an
artificial SINFONI data set constructed from an N-body/SPH simulation
(blue crosses). Both observations and simulation follow similar
distributions.} 
\label{fig:SBvsPos_comp} 
\end{figure}

The areas of intrinsically high star formation intensity are
blended after smearing the data, which lowers their intensities
(Fig.~\ref{fig:analogs}). Whether or not such peaks in the star-formation
intensity can be observed in principle depends on their relative
distribution and peak intensity relative to their surroundings within one
or two PSFs (cf. Figs.~\ref{fig:simus} and \ref{fig:analogs}). However,
Fig.~\ref{fig:SBvsPos_comp} demonstrates that, after smearing, the
simulated final radial distribution of surface brightness is consistent
with what we observe. The similarity seen does not depend on the
underlying velocity dispersion distribution assumed in the construction
of the synthetic data.

The star-formation intensities from the simulation decline by $\lsim$25\%
after beam smearing. While the precise value of this decrease is likely
to depend on the initial spatial distribution of the surface brightness,
it is always going to be small. The total star-formation rate, which
in the simulations is typically about 60 M$_{\sun}$ yr$^{-1}$, remains
essentially unchanged after making an artificial SINFONI data set. The
largest impact of smearing is on the distribution of the star-formation
intensities and in losing the lowest surface brightness emission because
of noise and the relatively low dynamic range of the data.

\subsection{Effect of beam smearing on velocity dispersions}

\begin{figure}
\setlength{\unitlength}{4.25cm}
\centering
\begin{picture}(2,1)
\put(0, 0){\includegraphics{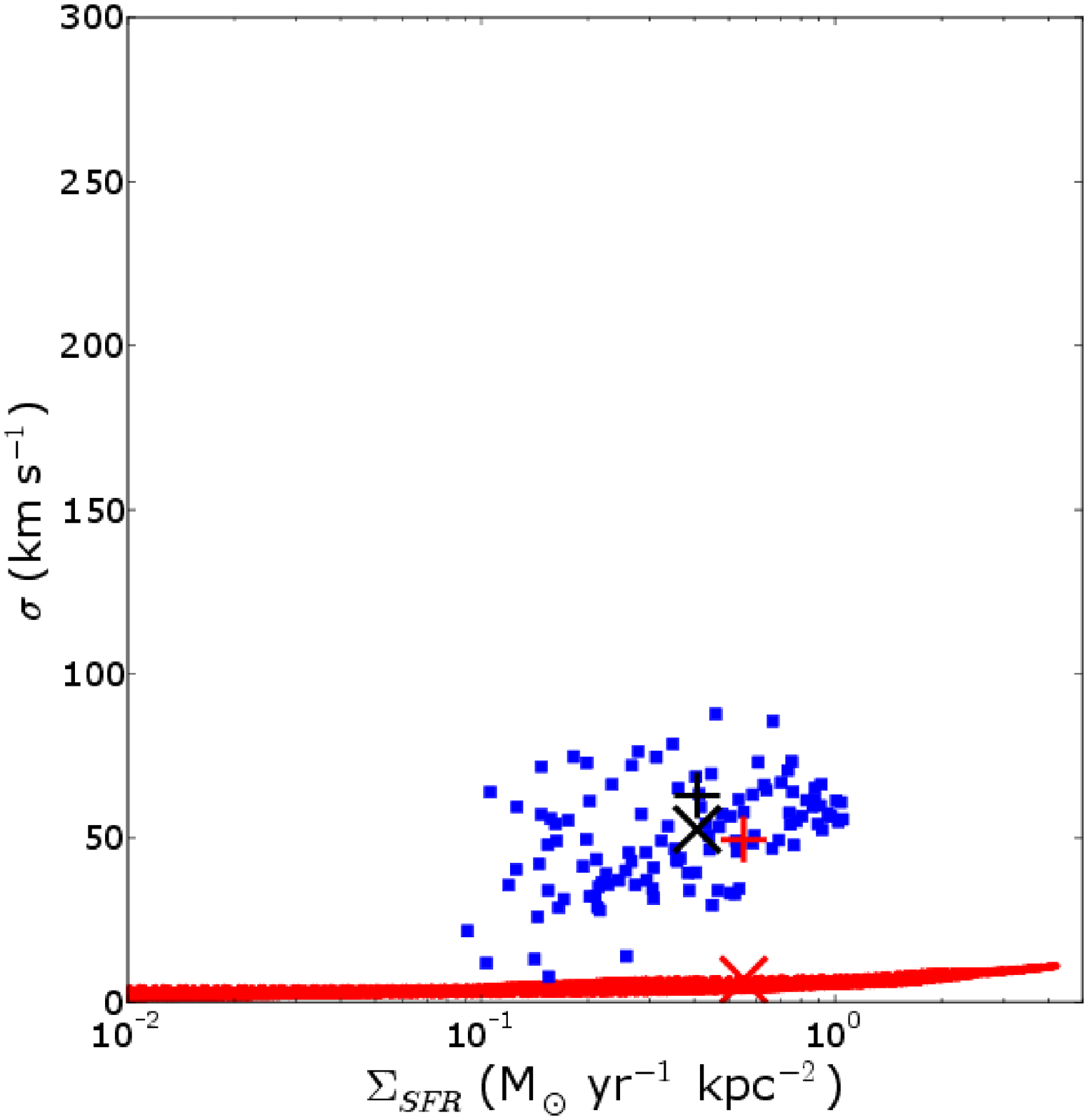}}
\put(1, 0){\includegraphics{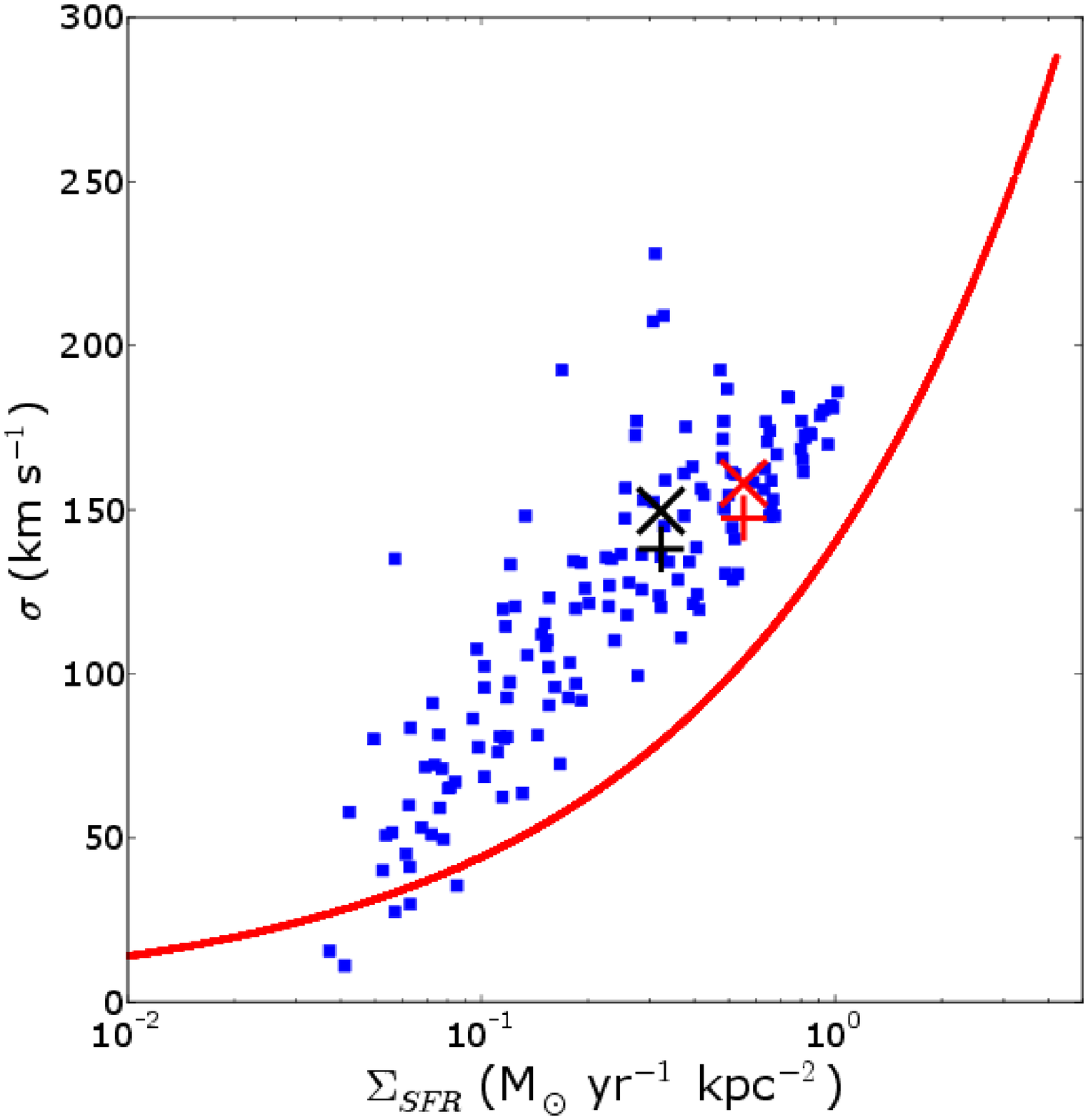}}
\end{picture}
\caption{In our analysis of the impact of beam smearing on the
velocity dispersion and star-formation intensity, we adopted two
relationshipsbetween these two quantities. One taken directly from
the simulations ({\it left panel}), and another where we assumed
$\sigma\propto\Sigma_{\rm SFR}^{1/2}$ ({\it right panel}). The model
velocity dispersions are shown in red (these are not lines but many
individual points). We also show the points (blue) after analyzing the
artificial data cubes derived from the simulation which include the
effects of beam smearing. The red $+$ (red $\times$) sign indicates the
mean star-formation intensity and integrated (original input) velocity
dispersion for the intrinsic relation between velocity dispersion
and star-formation intensity, while the black $+$ (black $\times$)
sign has the same meaning but now for the beam-smeared relationship.}
\label{fig:comp} 
\end{figure}

Since we have assumed two quite different relationships between velocity
dispersions and star-formation intensity, it is in the resulting velocity
dispersions that we see the most dramatic differences. In the model
with $\sigma_{\rm sim}$, the dispersion is low (a few to 10 km s$^{-1}$)
and almost constant as a function of radius and star-formation intensity
(Fig.~\ref{fig:comp}, left panel). In the synthetic observations however,
the beam smearing by the PSF (FWHM=\as{0}{6}) and the low spectral
resolution of the data broadens all lines by about $\sim$40 $\kms$
and the noise in the data increases the scatter about this mean rise
in the dispersions. Even though we correct for the intrinsic resolution
of the spectrograph, at relatively low S/N we tend to over-estimate the
widths of lines that are only marginally resolved. Our results suggest
that for galaxies with roughly constant and low dispersions ($\sim$40
km s$^{-1}$), a spectrograph like SINFONI will tend to over-estimate
the intrinsic widths of the lines but still produce a roughly constant
trend of dispersion with star-formation intensity.

For the model where the simulation dispersions are scaled as
$\sigma=140\sqrt{}\Sigma_{\rm SFR}$ km s$^{-1}$, there are almost no
differences between the mean velocity dispersions from the simulation
itself and the artificial data sets (Fig.~\ref{fig:comp}, right
panel). Because of the large range of the scaled velocity dispersions,
there is little effect of beam smearing on the average velocity
dispersion. Looking at the pixel-by-pixel quantities, we obviously
lose the relatively rare regions and peaks with the highest velocity
dispersions (over 200 $\kms$). However, as suggested by the average and
intrinsic dispersions, the decrease in the values (ranges and averages)
is not large, even after the smoothing to low spatial and relatively low
spectral resolution. Again, because of the loss of the lowest surface
brightness regions, we also tend to lose information on the areas with
the lowest velocity dispersions. However, in spite of all of this, the
general trend for higher star-formation intensity to result in higher
dispersion is maintained.

We also halved the signal-to-noise ratio and used the intrinsic
rotation speed of the simulated galaxy keeping the inclination angle at
45$\deg$. In both cases, the overall trend in the data is maintained and
the increase in the scatter in the star-formation intensity-dispersion
plane is not significant.

\subsection{Are the dispersions related to H$\alpha$ surface brightness?}

\begin{figure}
\includegraphics[width=9.0cm]{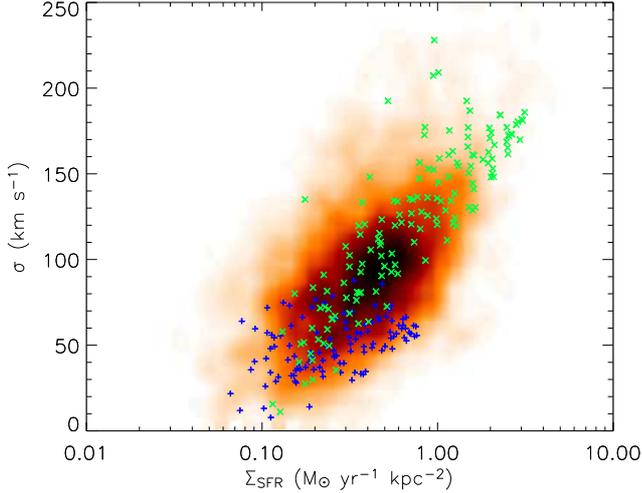}
\caption{Normalized 2-dimensional histogram of the frequency of occurrence
in the data set of \Ha line widths as function of star formation rate
intensity. The set of points for each galaxy are normalized such that all
of the galaxies have the same mean star-formation rate intensity (their
distribution of the velocity dispersion in each galaxy is unaffected). The
mean value was chosen to match the mean value of the star-formation
intensity in the model with $\sigma=(\epsilon\Sigma_{\rm SFR})^{1/2}$ as
indicated by the red $+$ sign in the right panel of Fig.~\ref{fig:comp}
($\sim$0.5 \Msunperyrperarea). As in Fig.~\ref{fig:SBvsPos}, the darkest
regions represent the highest frequency of occurrence in the data set. The
green $\times$ and blue $+$ symbols represent our simulated data after
beam smearing, assuming that the two quantities plotted are related as
$\sigma=(\epsilon\Sigma_{\rm SFR})^{1/2}$ and $\sigma$=$\sigma_{\rm sim}$,
respectively.} 
\label{fig:blend} 
\end{figure}

We now compare the observed trend between star-formation intensity and
\Ha line velocity dispersion with the models we used to make synthetic
data sets, including the effects of beam smearing. Fig.~\ref{fig:blend}
shows that the values of the simulated \Ha line dispersion $\sigma_{\rm
sim}$ are too low to explain the large line widths we generally observe
in our sample of galaxies. On the other hand, the simulation using the
relationship, $\sigma\propto\Sigma_{\rm SFR}^{1/2}$, is a good match to
the overall distribution of the data.  We note that more properly,
a constant term should perhaps be included in the relationship, as
$\sigma\propto\Sigma_{\rm SFR}^{1/2}$ + constant,  which is both warranted
by observations and expected theoretically \citep{dib06}.  At star
formation intensities of order 10$^{-4}$ to 10$^{-5}$ M$_{\sun}$ yr$^{-1}$
kpc$^{-2}$, the observed velocity dispersion of the H$\alpha$ emission
line are $\sim$5-20 km s$^{-1}$ \citep{faithi07, erroz-ferrer12}. We
have chosen to ignore this constant in our analysis for two reasons.
First, velocities this small are below the resolution of the SINFONI
data. Thus we are unable to probe the characteristics of galaxies with
low star formation intensities.  Second, the star formation intensities
of the galaxies we are observing are several orders of magnitude greater
and this including a small constant term would have an insignificant
impact on our analysis or conclusions.

\begin{figure*}
\setlength{\unitlength}{9.0cm}
\begin{picture}(2,1)
\put(1, 0){\includegraphics{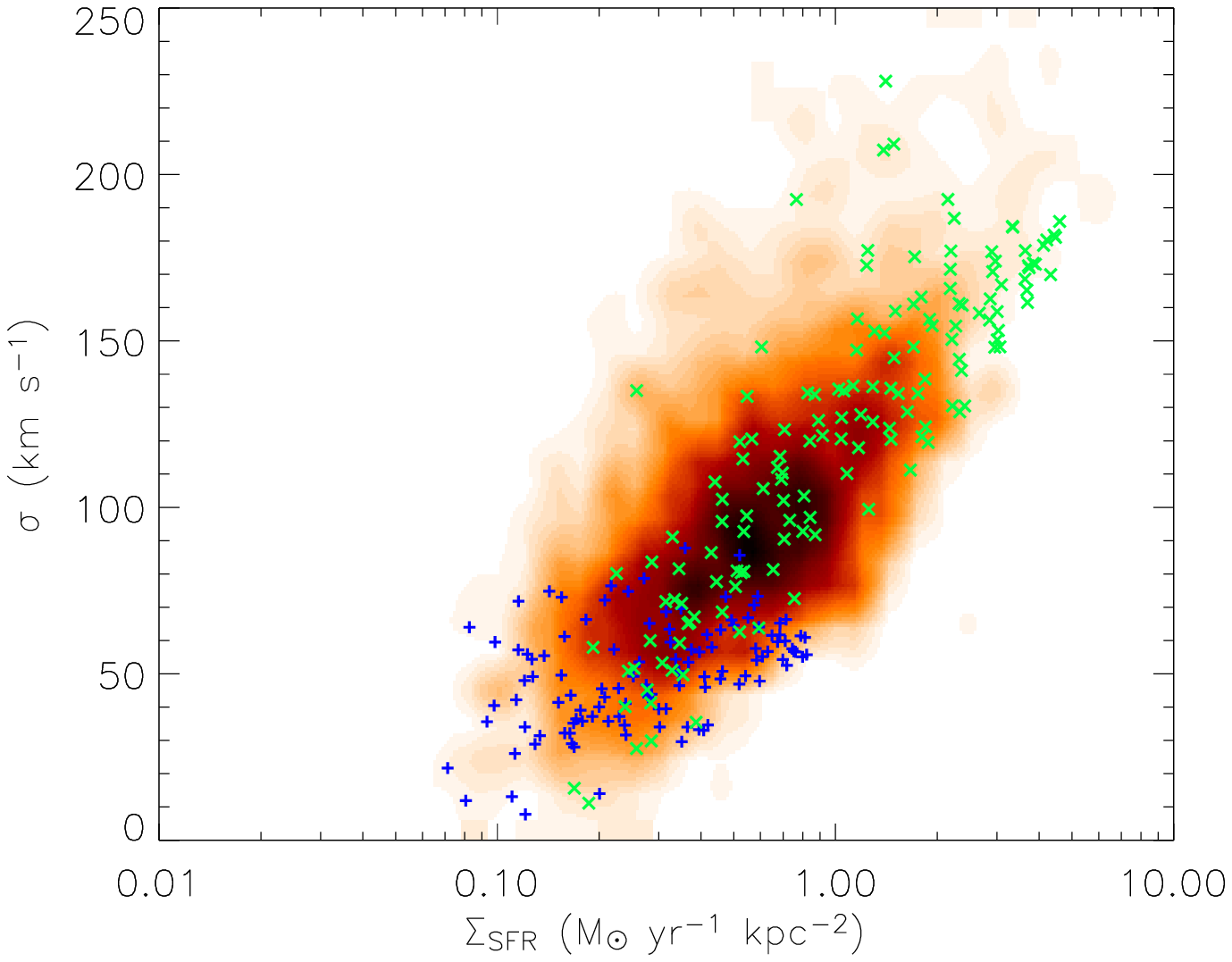}}
\put(0, 0){\includegraphics{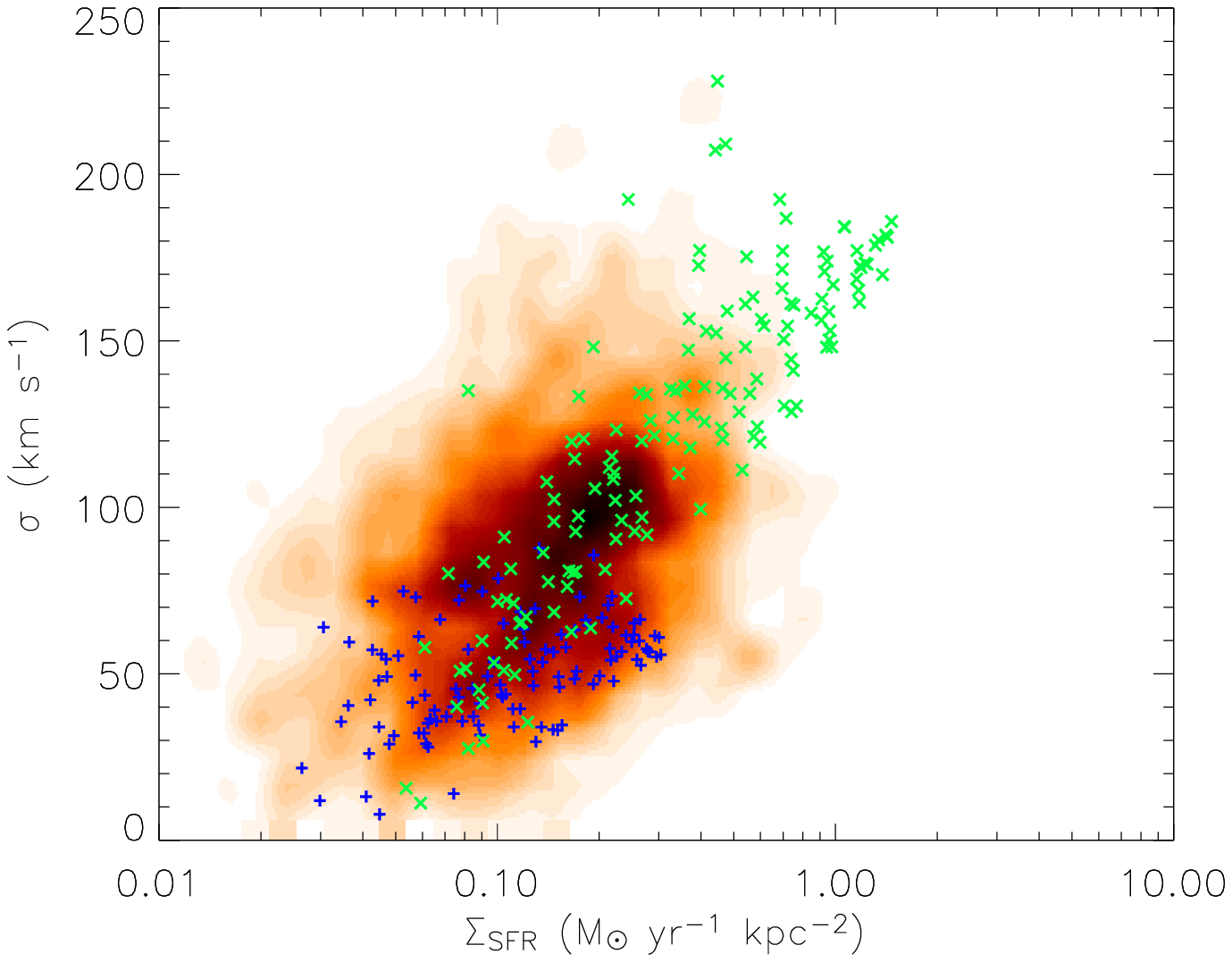}}
\end{picture}
\caption{Plots of the normalized frequency of occurrence of H$\alpha$
velocity dispersion versus star-formation intensity in two redshift bins,
z$<$1.8 ({\it left}) and z$>$2 ({\it right}) (cf. Fig.~\ref{fig:blend}
for the whole sample). The normalizations are different from that
used in Fig.~\ref{fig:blend} and in the left and right panels. For
both sub-samples, the average star-formation intensity of each
galaxy was shifted to the average of its ensemble of galaxies
(0.16 \Msunperyrperarea for z$<$1.8 and 0.73 \Msunperyrperarea
and z$>$2). Note that the two sub-samples tend to sample different
ranges in both velocity dispersion and star-formation intensity. The
green $\times$ and blue $+$ symbols represent our simulated data
after beam smearing, assuming that the two quantities plotted are
related as $\sigma=(\epsilon\Sigma_{\rm SFR})^{1/2}$ km s$^{-1}$ and
$\sigma$=$\sigma_{\rm sim}$, respectively. The green and blue model
points have been shifted to overlap with the data (the location of the
modeled points along the axis of $\Sigma_{\rm SFR}$ is set by the coupling
efficiency of the mechanical energy output of young stars to the ISM).}
\label{fig:Zevol}
\end{figure*}

Figure~\ref{fig:Zevol} displays the differences in the H$\alpha$
velocity as a function of the star-formation intensity distribution for
the galaxies in two different redshift regimes, z$<$1.8 and z$>$2. The
key result is that the data are consistent with a simple scaling
between dispersion and star-formation intensity in both samples --
however, on average the z$\sim$1.5 galaxies have lower dispersions and
star-formation intensities. Also, the scatter is larger for the lower
redshift galaxies, which perhaps indicates that a single relationship of
the type $\sigma\propto\sqrt{}\Sigma_{\rm SFR}$ does not explain all of
the trends in the lower-redshift data as well as it does for the higher
redshift objects.

At the low end of the star-formation intensity within individual galaxies
and over the ensemble of galaxies with lower average star-formation
intensities, it is likely that we are over-estimating the velocity
dispersions, and it could be that the dispersions are becoming roughly
constant with radius. We say this, because our artificial model-based
data cubes suggest that observed dispersions smaller than $\sim$40-50
\kms\ are consistent with $\lsim$10 \kms, a level at which we simply
can no longer differentiate between a model where the line widths are
proportional to $\Sigma_{\rm SFR}^{1/2}$ or one where the line widths
are consistent with what is observed in the MW and other nearby galaxies
\citep[see, e.g.][]{wada02,dib06,agertz09}.

In Figures~\ref{fig:Zevol}, we plotted the simulated data against
the normalized distribution of star formation intensity and velocity
dispersion. However, this does not give a sense of the differences between
galaxies at different redshifts in their distribution of star formation
intensity and velocity dispersion. The distributions of the galaxies
in the higher redshift regime are consistent with those of 11 galaxies
presented in \citet{L09}.  Quantitatively, the ensemble of galaxies with
z$>$1.8 and z$<$1.8 have a mean star formation intensity of 0.73$\pm$0.09
and 0.16$\pm$0.02 M$_{\sun}$ yr$^{-1}$ kpc$^{-2}$ respectively.  Since
both samples contain about the same number of galaxies, the width of the
distribution of the higher redshift sample is significantly larger and the
difference in width between the samples is consistent with \citet{L09}.

Moreover, correcting individual galaxies in the sample for the average
extinction \citep[derived from fitting the spectral energy distribution
with stellar population synthesis models;][]{fs11} does not reduce the
scatter among them.

\subsection{The equivalence of integrated and spatially resolved
measurements}

So far, we have studied the spatially resolved relationship between the
\Ha velocity dispersion and the star-formation intensity as probed by the
\Ha emission. In \citet{LT11a}, we argued that the relationship between
the integrated \Ha luminosity (or equivalently the total star-formation
rate) and surface brightness-weighted mean dispersion of whole galaxies
reveals the same underlying physical mechanism as the spatially resolved
measurements -- namely, that the intense star-formation is responsible
for the dynamics of the warm ionized medium as probed through the \Ha
recombination line \citep[see also][]{Green10}.

To further investigate this relationship, we have tested the idea
that the intense star-formation in these galaxies can even explain
the integrated measurements of the total star-formation rate and the
integrated velocity dispersion. This is important: if we can show that
also these integrated measurements reveal this underlying relationship,
then a greater wealth of data become available for analysis, since slit
spectra are more commonly available than integral field spectra.

We first note that we find little difference in velocity dispersion for
whole galaxies if we simply sum up the \Ha emission pixel-by-pixel and
estimate their integrated dispersion or if we weigh each measurement
by surface brightness (or signal-to-noise). The difference between both
methods is only about 15 km s$^{-1}$, where flux weighting tends to give
a lower value and a scatter of about 15 km s$^{-1}$. So compared to the
mean integrated velocity dispersion of our sample, this represents a
systematic offset of only about 10\% \citep{LT11a}.

\begin{figure}
\includegraphics[width=9.0cm]{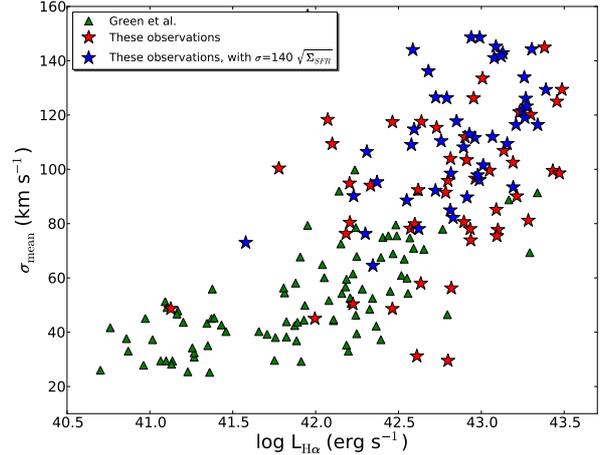}
\caption{Mean H$\alpha$ line velocity dispersion, $\sigma_{\rm mean}$
(in \kms), versus total \Ha\ luminosity (in erg s$^{-1}$) for different
samples of local and high redshift galaxies. Local galaxies (green
triangles) are taken from \citet{Green10} and the model results for
our sample are represented by red and blue stars. The blue stars show
the results where we assign to each pixel in a galaxy an \Ha surface
brightness based on the model relationship, $\sigma=140\sqrt{}\Sigma_{\rm
SFR}$ km s$^{-1}$ and use the observed local \Ha velocity dispersion. We
then estimate the total \Ha luminosity by summing up the predicted
pixel-by-pixel surface brightnesses, and estimate $\sigma_{\rm mean}$
by weighting the pixel-by-pixel velocity dispersion measurements by the
\Ha\ surface brightnesses derived from the model.}
\label{fig:ha_vs_sigmean}
\end{figure}

To further test the relevance of a relationship of the form,
$\sigma$=($\epsilon \Sigma_{\rm SFR})^{1/2}$, where $\epsilon$ is the
efficiency at which the mechanical energy from a star-formation intensity
$\Sigma_{\rm SFR}$ is converted into turbulence and bulk flows in the
interstellar medium, we applied it to the spatially resolved dispersion
measurements, to estimate integrated quantities. Specifically, we used
the spatially resolved, pixel-by-pixel \Ha line velocity dispersion
measurements to estimate the \Ha surface brightness of each pixel. We
then used these surface brightness estimates as weights when combining
the individual pixel-by-pixel velocity dispersions, to estimate the
integrated mean dispersion in H$\alpha$, $\sigma_{\rm mean}$. We summed
these estimated surface brightness values to estimate the integrated \Ha
fluxes and then converted these to total luminosities using our adopted
cosmology. The results of this analysis show that we can reproduce
the actual measurements reasonably well considering the crudeness
of the model and that we only considered one coupling efficiency
(Fig.~\ref{fig:ha_vs_sigmean}). So not only does this simple scaling
relationship reproduce the spatially resolved measurements, but it
can also reasonably reproduce the integrated measurements. Apparently,
unlike in many models of the ISM, the observed velocity dispersion of
the warm ionized medium does increase with increasing star-formation
intensity and rate \citep{dib06, ostriker11}.

\section{Discussion}\label{sec:discussion} 

\begin{figure}[h!]
\setlength{\unitlength}{9.0cm}
\includegraphics[width=9.0cm]{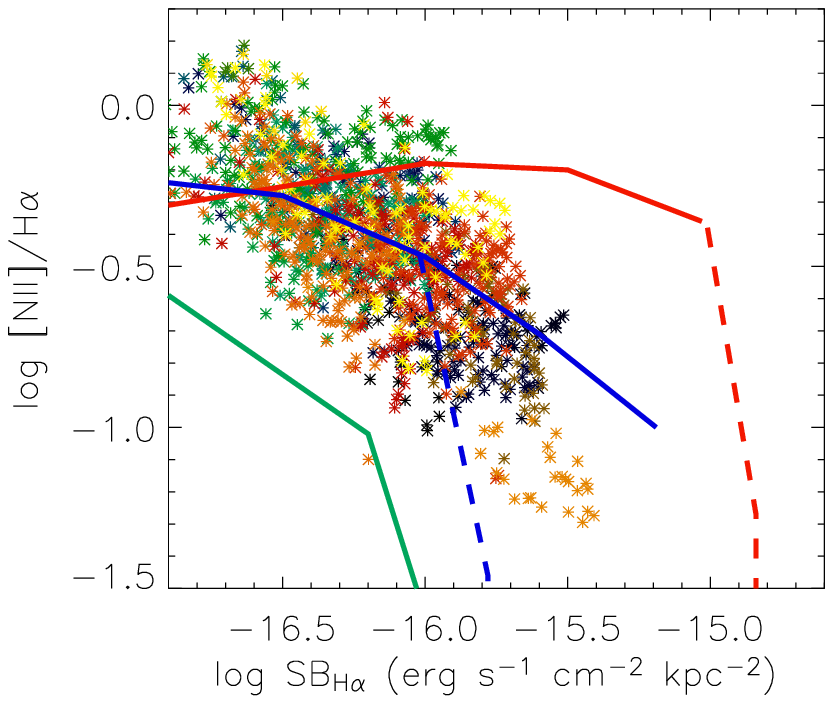}
\includegraphics[width=9.0cm]{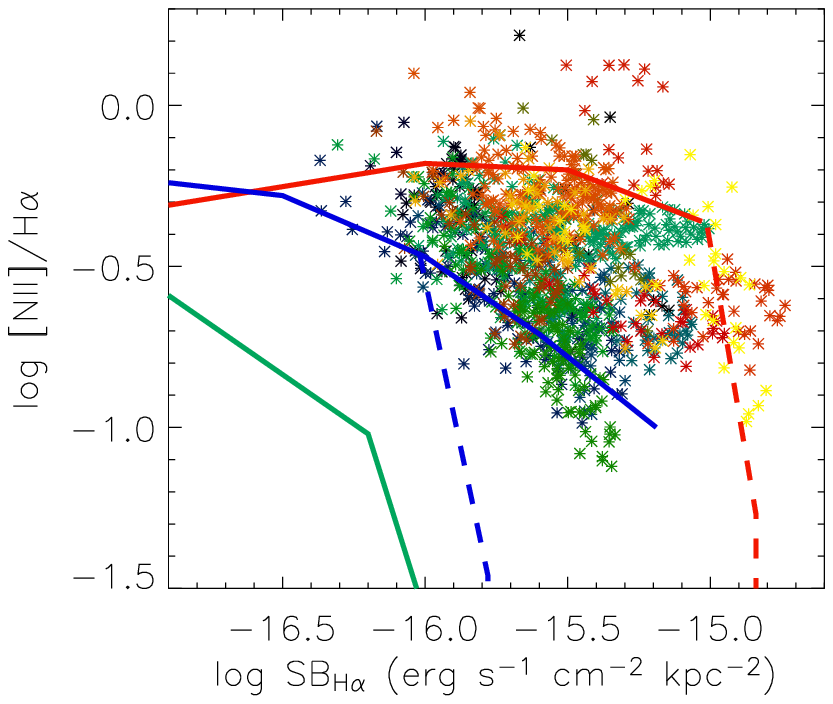}
\caption{H$\alpha$ surface brightness (corrected for cosmological
surface brightness dimming) versus the logarithm of the [N{\sc
ii}]$\lambda$6583/\Ha line ratio. The galaxies have been separated into
two redshift bins: z$<$1.8 ({\it top}) and z$>$2 ({\it bottom}). The
lines represent results for photo-ionization modeling (see text for
details) for 6 sets of conditions: a range of hydrogen densities,
log n$_H$ (cm$^{-3}$) =1 (in green), 2 (blue) and 3 (red), and
of column densities, log N$_H$ (cm$^{-2}$)=21 (solid lines) and 20
(dashed lines). The ionization parameters range from log U=$-$5 to $-$1
(increasing from left to right along the lines, meaning low ionization
parameters have relatively high log [N{\sc ii}]/H$\alpha$ and low
H$\alpha$ surface brightnesses). At low ionization parameters the solid
and dashed lines for the column densities overlap until the lower column
density cloud becomes density bounded, at which point the models reach
very low ratios of [N{\sc ii}]$\lambda$6583/H$\alpha$ (as shown by the
sharply dropping dotted lines). For clarity, we only show the log N$_H$
(cm$^{-2}$)=21 and log n$_H$ (cm$^{-3}$) =3 line until log U = $-$3 and
for log N$_H$ (cm$^{-2}$)=21 and log n$_H$ (cm$^{-3}$) =2 until log U =
$-$2.} 
\label{fig:pressure} 
\end{figure}

Figure~\ref{fig:pressure} shows that [NII]/H$\alpha$ ratios are
not arbitrary, but scale with H$\alpha$ surface brightness. This has
already been noted by \citet{L09} and is confirmed with our new, larger
sample. The emission line ratios and \Ha surface brightnesses are related
to the gas pressure, column density, intensity of the radiation field
and the number of line-emitting regions along the line of sight within
each beam. New, compared to our previous analysis of this relationship,
is that our expanded sample includes galaxies at lower redshift, which
in turn will allow us to probe pressures, column densities and radiation
fields in galaxies with less extreme \Ha surface brightnesses.

Measurements of the \SII\ doublet ratio in individual galaxies in a
sub-sample of this sample \citep{L09} and through a stacking analysis
of this sample \citep{LT11b}, suggest that the typical densities of
the warm ionized gas are  n$_{\rm e}$$\sim$10s to $\sim$500 cm$^{-3}$.
To compare these estimates, we ran a suite of photoionization models
using version 08.01 of the Cloudy code \citep{ferland98} to show the
relationship between the [N{\sc ii}]$\lambda$6583/\Ha line ratio and the
\Ha surface brightness as a function of the hydrogen internal pressure,
column density and ionization parameters for a single ionized cloud
\cite[see][]{L09}. Unfortunately, the interpretation of such modeling is
ambiguous due to the degeneracy between surface brightness, line ratio and
the number of individual emission line regions per large physical beam of
our observations. Given the crude sampling of our data, it is likely that
the densities are lower than predicted by a single cloud model, as would
be the ionization parameters. Therefore, for both the higher and lower
redshift data sets, the likely densities are certainly $<$1000 cm$^{-3}$
but generally larger than a few times 10 cm$^{-3}$ and are consistent
with estimates of the electron density directly measured from the \SII\
doublet ratio in individual galaxies and a stacking analysis.

There is an interesting difference in the regions occupied in the
[N{\sc ii}]$\lambda$6583/\Ha versus \Ha surface brightness plane by
galaxies at z$<$1.8 and z$>$2. Taken at face value, it would suggest
we are probing less dense gas in the lower redshift galaxies. However,
this is undoubtedly due to probing lower surface brightness gas, related
to the strong influence of surface brightness dimming, which allows us
to probe the very diffuse, perhaps even shock heated, gas in the lower
redshift sample (the regions with relatively low surface brightness and
high ratios of [N{\sc ii}]$\lambda$6583/\Ha in Fig.~\ref{fig:pressure}
likely require a contribution to their ionization from shocks). This
effect is observational, rather than evolutionary. A complete census
of both redshift regimes would be necessary to show that the average
pressure or number of overlapping emission line regions (i.e. thickness)
in the warm ionized media of galaxies is declining with redshift.

The thermal pressures estimated from the electron densities are much
higher than in the disk of our Milky Way or other nearby normal
galaxies. The hydrostatic mid-plane pressure in the MW is about
10$^{3.3-4.3}$ K cm$^{-3}$, while for other local spirals it can amount
to about 10$^{6}$ K cm$^{-3}$, though it is typically about 10$^{4.6}$
K cm$^{-3}$ \citep[][and references therein]{blitz06}. In the nuclear
regions of local starburst galaxies, thermal pressures estimated from
the \SII\ doublet ratio are typically about 10$^{6-7}$ K cm$^{-3}$
\citep{lehnert96}. The pressures we estimated from the photo-ionization
models are consistent with those of nearby intensely star-forming galaxies
\citep{heckman90, lehnert96, wang98} and measurements of density-sensitive
lines at high redshift \citep[\SII\ doublet ratio;][]{LT11b, L09,
nesvadba07}.

By comparing these thermal pressures in the WIM to that generated by
the radiation pressure of the ionizing radiation field, it is possible
to constrain the general distribution of the WIM \citep{krumholz09} .
Our optical line ratios suggest that the gas is predominately
photoionized by massive stars \citep{L09}. To estimate the impact
of radiation pressure on an ionized fluid, it is appropriate to only
consider the ionizing radiation. We can use the ionization parameter to
gauge the relative contribution of radiation pressure to  the thermal
pressure of the WIM. Simple scaling of the definition of the ionization
parameter suggests P$_{\rm rad}$/P$_{\rm thermal}$= 2.5$\times$10$^{-2}$
U ($<$h$\nu$$>$/18 eV)/k(T/10$^4$ K) for T=10$^4$ K and $<$h$\nu$$>$=18
eV, where $<$h$\nu$$>$ is the average energy of ionizing photons and k
is the Boltzmann constant. Since the ionizing parameter is always less
than log U = $-$1.6 (and mostly $\la -$3), the radiation pressure is
lower than the thermal gas pressure throughout the nebulae. Finding
that the thermal pressure dominates the radiation pressure and that
the ionization parameters are relatively low, suggests that the WIM is
widely distributed over the volume of the ISM in our sample of galaxies,
truly part of the general interstellar medium, and not simply near or
directly associated with the star forming clumps or regions \citep[see
][ for a detailed discussion of this point]{krumholz09}. This point is
crucial for our subsequent analysis as it suggests that the WIM is a
tracer of the dynamics of the general ISM in distant galaxies and not
merely  of the \HII\ regions or clumps.

Other models or relationships have been proposed to explain the
high dispersion in the warm ionized gas that are directly related to the
ionization state of the gas. \citet{wisnioski12} and \citet{swinbank12}
propose that the relationship between star formation intensity
and velocity dispersion can be explained by a relation between
the luminosity of a Hydrogen recombination line (e.g. H$\beta$)
and the gas velocity dispersion, as observed in local \HII\ regions
\citep[e.g.][]{terlevich81}.  The sizes of local \HII\ regions and
the relationship between line luminosity and velocity dispersion can
be understood simply if the gas within the \HII\ regions is virialized
\citep{terlevich81}.  In such a hypothesis, the line widths are mainly
due to virial motions, implies that the warm ionized medium is in
close proximity to the star forming clumps (within the gravitational
sphere of influence).  Nearby HII regions, for which such a relation
is appropriate, generally show ionization parameters that are high, log
U$\approx -$1 to $-$2\citep[e.g.][]{lopez13}, but our analysis suggests
that distant galaxies generally have much lower ionization parameters,
implying that their emission line gas luminosity is dominated by more
widely distributed gas and not directly associated with the clumps.

\subsection{Source of the pressure in the WIM} 
\label{pressuresource}

What are the sources of pressure in these distant galaxies?  Is the
pressure generated by the thermalization of the collective mechanical
energy output of massive stars through the thermalization of stellar
winds and supernova explosions? Observations of nearby galaxies suggest
that the low ionization line emitting region is approximately in
pressure equilibrium with the X-ray emitting plasma, with thermal
pressures in the range 10$^{-9}$ to 10$^{-10}$ dyne cm$^{-2}$
\cite[e.g.][]{heckman90,moran99,westmoquette07, strickland09}. Similar
pressures are found in models using typical initial conditions expected
for galaxies exhibiting outflows \citep[e.g.][]{suchkov96}. If these
two phases are in equilibrium then we can use a simple model to estimate
the pressures in the hot X-ray emitting plasma to see if it agrees with
our estimates of the thermal pressures (densities) in the warm ionized
media in distant galaxies.

\citet{strickland09} has shown that the simple analytic expressions
of \citet{chevalier85} for the pressure, temperature and velocity of
outflows driven by the intense energy injection from young massive
stars agree with hydrodynamical simulations. Adopting and scaling
these simple relations from \citet{chevalier85} we estimate the central
pressure in a starburst region as P$_{\rm c}$= 3.5$\times$10$^{-10}$
($\zeta$$\dot{\rm M}_{0.26}$)$^{1/2}$($\epsilon_{\rm thermal}$$\dot{\rm
E}_{\rm 41.9}$)$^{1/2}$ R$_{\rm SB, 350}^2$ $\Omega$ dyne cm$^{-2}$,
where $\zeta$ is the mass-loading factor, $\dot{\rm M}_{\rm 0.26}$ is the
mass return rate in units of 0.26 M$_{\sun}$ yr$^{-1}$, $\epsilon_{\rm
thermal}$ is the thermalization efficiency of the stellar winds and
supernovae, $\dot{\rm E}_{\rm 41.9}$ is the energy injection rate in units
of 7.4$\times$10$^{41}$ erg s$^{-1}$, R$_{\rm SB, 350}$ is the radius of
the starburst in units of 350 pc and $\Omega$ is the opening angle of the
outflow in sr. The adopted scalings for the mass and energy injection
rates from massive stars are appropriate for a star formation rate of
1 M$_{\sun}$ yr$^{-1}$ \citep{leitherer99} within a radius consistent
with clump sizes and \Ha disk thicknesses in high redshift galaxies
\citep{fs11,epinat12}. For the radius scaling, see \citet {strickland09}.

\begin{figure}[ht!]
\includegraphics[width=9.0cm]{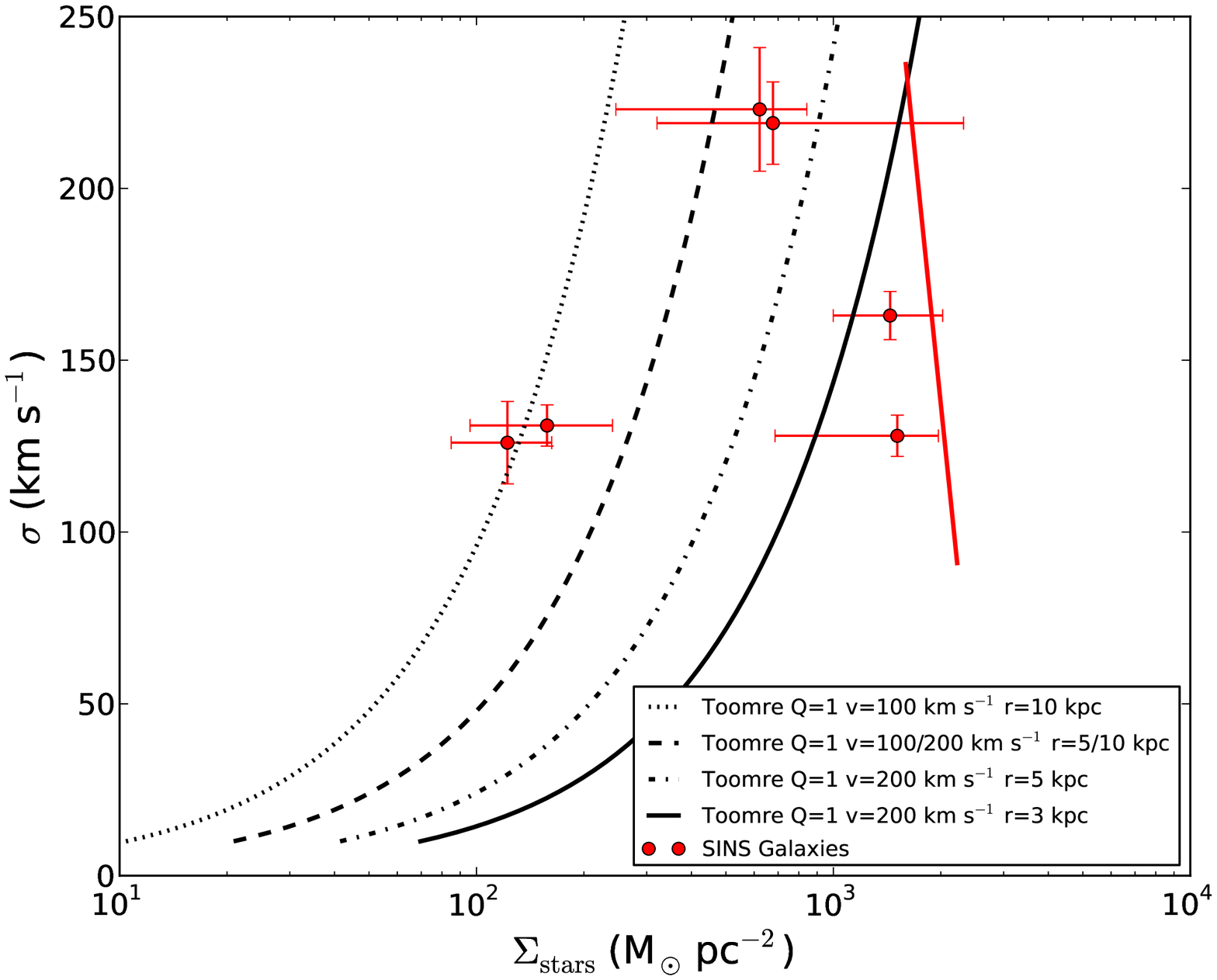}
\caption{\Ha velocity dispersion, $\sigma$ as function of stellar mass
surface density, $\Sigma_{\rm stars}$ (M$_{\sun}$ pc$^{-2}$). The lines
represent the Toomre criterion Q=1 for two rotation speeds (v=100 or
200 km s$^{-1}$) and three radii (r=3, 5, 10 kpc), as indicated in the
legend. These values span the range of radii and velocity shears observed
in these galaxies \citep{fs06, fs09}. The stellar mass surface densities
(red dots) are estimates within one effective radius for 6 galaxies
in our sample with NICMOS images (see text). The red solid line shows
the relationship between central velocity dispersion and stellar mass
surface density for local early-type galaxies \citep{shen03, taylor10}.}
\label{fig:toomreQ}
\end{figure}

Reasonable assumptions for the thermalization efficiency, mass loading
factor, and mass and energy injection rates, result in pressures of
10$^{-9}$ to 10$^{-10}$ dyne cm$^{-2}$ in our galaxies, similar to
what is estimated for nearby starburst galaxies. This simple modeling
suggests that the hot plasma generated by the thermalization of the
mechanical energy output of young stars can generate sufficient pressure
to approximately match our estimates for the thermal pressure in the
warm ionized medium.

One of the assumptions that is needed for the model to be appropriate is
that the energy injection rate is approximately constant over an outflow
timescale \citep[e.g.][]{chevalier85, suchkov96,strickland09}. This has
been shown to be true for local starbursts \citep[e.g.][]{heckman90,
lehnert99}. While we do not know the outflow timescale of the distant
galaxies, since their star-formation intensities are similar to nearby
starburst galaxies \citep[see discussion in][]{L09}, it is logical to
assume their outflow timescales would be similar (10s Myr or less).
Distant galaxies similar to the ones observed here \citep[e.g.][]{erb06,
fs11} have star formation timescales of $\approx$100s Myrs and are
likely much longer than the outflow timescales. A reasonable assumption
is that the energy injection rate is approximately constant in distant
galaxies as it appears to be in nearby starbursts and thus the model of
\citet{chevalier85} is an appropriate approximation for estimating the
pressures \citep[see also ][ for a discussion]{strickland09}.

\subsection{Mass and energy flow?}\label{picture} 

The aforementioned rough equality of the thermal pressure in the two ISM
phases, the hot 10$^7$--10$^8$ K plasma and the 10$^4$ K warm ionized
medium, is important. It may imply that the mechanical energy injected
into the ISM of the galaxy couples efficiently with the recombining gas
and perhaps drive a mass and energy flow from the WIM to the cold neutral
and molecular medium \citep{guillard09, guillard10}. The recombining gas
is unstable at the pressures we observe and will quickly cool, collapse
and become part of the cold neutral medium, CNM \citep{wolfire95}. Since
this is expected to happen quickly this would imply that the CNM acquires
the relative kinematics of the WIM \citep{guillard12}. If the gas is
dusty, this recombining gas will rapidly turn molecular since the H$_2$
formation time scales on dust grains is much shorter than the expected
dynamical time scales of the recombining, cooling and collapsing clouds
\citep{guillard09}. The formation of gravitationally bound clouds releases
gravitational energy as turbulence into the molecular gas.

This picture links the energy input from intense star formation to energy
dissipation, the formation of molecular gas and star formation. Thus not
only is this a flow of mass and energy  but also one of kinetic energy,
which will not only contribute to the turbulent energy of the molecular
gas but also serve to balance dissipation. Of course, it is not a simple
one-directional flow, but a cycle or feedback loops where individual
molecular clouds may form stars when they become self gravitating, or
simply be heated and destroyed. Naturally, because all the timescales
in the cycle are relatively short, $\la$10 Myrs, the ISM will be in a
dynamical equilibrium. The components of the ISM should not be viewed
as single, isolated components but rather as an intricate balance whose
relative mass flow rates are determined by a competition between the
heating and cooling rates as well as the outflow rate due to starburst
driven winds and the infall rate due to the cosmological accretion of
gas through cooling gas and mergers.

One of the conditions that must be met in such a picture is that the
mass deposition rate of the WIM should be more than sufficient to
fuel star formation, or $\dot{\rm M}_{\rm SF}\la\epsilon_{\rm star
form} \dot{\rm M}_{\rm HII}$, where $\epsilon_{\rm star form}$ is
the efficiency of star formation. The recombination rate of the \Ha\
emitting gas is given by simple ionization balance, $\dot{\rm M}_{\rm
HII}$=1.9$\times$10$^5$ L$_{\Ha , 43}$ M$_{\sun}$ yr$^{-1}$. The star
formation rates in these galaxies are $\sim$10-300 M$_{\sun}$ yr$^{-1}$
and thus the WIM recombination rates can supply the gas for star formation
as long as the star formation efficiency is $\ga$0.1\%.  For a reasonable
$\epsilon_{\rm star form}$ only a small fraction, a few  to about 10
percent, of the recombining gas  would then be necessary to fuel this
cycle. So at least in terms of mass flow rates, there is sufficient mass
in the warm ionized medium to plausibly sustain such a cycle.

Of course, for the energy injection to be effective in regulating
star formation, as we have hypothesized, it should drive the galaxy
towards the line of Toomre disk stability, Q$\sim$1. This occurs if
the star formation can inject sufficient turbulence to roughly balance
gravitational collapse and hydrostatic pressure.

\begin{figure}[hb!]
\includegraphics[width=9.0cm]{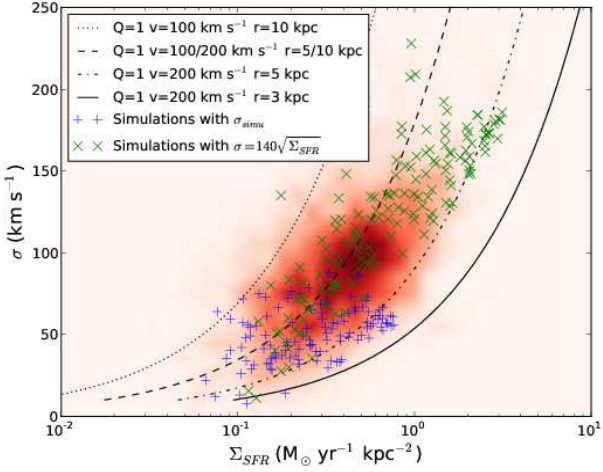}
\caption{\Ha velocity dispersion, $\sigma$, as function of the
star-formation intensity, $\Sigma_{\rm SFR}$ (\Msunperyrperarea). The
lines represent the Toomre criterion for gas, assuming both Q=1 and
that the Schmidt-Kennicutt relation is appropriate for a variety of
rotation speeds and radii, as in Fig.~\ref{fig:toomreQ} and as indicated
in the legend (see text for details). The underlying display of the
normalized data (shown in various shades of red) is the same as 
in Fig.~\ref{fig:blend} (see that figure and text for details).}
\label{fig:toomreQSFR}
\end{figure}

While the WIM traces the overall dynamics of the ISM (see beginning
of \S~\ref{sec:discussion}), its contribution to the turbulent
pressure is negligible. It is given by ${\rm P}_{\rm gas, turb}$ =
$<\rho_{\rm gas}>\sigma_{\rm gas}^{2}$ where $<\rho_{\rm gas}>$ is the
volumeweighted mean density (=$\rho_{\rm phase} f_{\rm V,phase}$, where
$\rho_{\rm phase}$ is the intrinsic density and $f_{\rm V,phase}$ the
volume filling factor of the ISM phase). The total turbulent pressure
is actually the sum of those of all individual ISM phases, but we have
suppressed the summation for clarity. The \Ha\ emitting gas likely does
not contribute significantly to the total turbulent pressure. Assuming
simple case B recombination, T$_e$=10$^4$ K and equality of proton and
electron densities, we can crudely calculate the emitting volume needed
to supply the \Ha\ luminosity compared to the volume over which it is
observed. The volume filling factor of the recombining gas is ff$_{\rm
V, WIM}\sim$V$_{\rm em}$/V$_{\rm n}$ = 4$\times$10$^{-5}$ L$_{\Ha ,43}$
n$_{\rm e, 300}^2$ V$_{\rm n, 100}^{-1}$, where V$_{\rm em}$ is the
emitting volume, V$_{n, 100}$ is the actual ISM volume in units of 100
kpc$^3$ (the typical isophotal area is 100 kpc$^2$ and we assumed a disk
thickness of 1 kpc), L$_{\rm \Ha ,43}$ is the \Ha\ luminosity in units
of 10$^{43}$ erg s$^{-1}$ and n$_{\rm e, 300}$ is the electron density
in units of 300 cm$^{-3}$. The filling factor of the recombining gas
is $\sim$10$^{-6}$.

The most significant contribution to the total turbulent pressure is
likely from the cold neutral and molecular medium. We can estimate
the density of the dense molecular gas using observations of distant
galaxies with star formation rates and general properties similar
to ours. H$_2$ gas mass surface density measurements are available
for galaxies similar to those studied here \citep[z$\sim$1 and
2;][]{daddi10,aravena10,dannerbauer09,tacconi10,combes11} which suggest
that at z$\sim$2 the mean value is about 500 M$_{\sun}$ pc$^{-2}$ and
that total gas fractions $\sim$30--50\% \citep{daddi10,tacconi10}. The
volume weighted H$_2$ gas density, $<n_{H_2}>$=$\Sigma_{\rm gas}$/H
m$_{\rm H_2}$ = 10.2 $\Sigma_{\rm gas, 500}$/H$_{\rm 1 kpc}$, where
$\Sigma_{\rm gas, 500}$ is the gas mass surface density in units of
500 M$_{\sun}$ pc$^{-2}$ and H$_{\rm 1 kpc}$ is the thickness of the
gas layer in units of 1 kpc. This implies that P$_{\rm turb, mol}$=3
$\times$10$^{-9}$ $<\rho_{\rm gas, 10}>$ $\sigma_{\rm H_2, 100}^{2}$,
where $<\rho_{\rm H_2, 10}>$ is the H$_2$ volume weighted particle density
in units of 10 cm$^{-3}$ and $\sigma_{\rm H_2, 100}$ is in units of 100
km s$^{-1}$. These estimates suggest that the turbulent pressure in the
molecular gas may be high \citep{swinbank11}.

For a galaxy to lie near the line of stability, Q=1,  the pressure due to
gravity and that due to turbulence should be roughly equal. To estimate
the pressure due to gravity we will assume hydrostatic equilibrium, which
is obviously an over-simplification. However, for the present analysis,
we are concerning ourselves with globally averaged estimates, suggesting
that hydrostatic equilibrium is a reasonable assumption. It is impossible
to get direct estimates for the hydrostatic gas pressures for our sample
of galaxies as stellar mass surface densities are available for only a
few objects  \citep{fs11}. As with the estimates of the volume-weighted
gas mass densities, we will have to rely on results from similar samples
of distant galaxies, which find stellar mass surface densities of 100-2000
M$_{\sun}$ pc$^{-2}$ \citep[e.g.][]{barden05,yuma11,fs11}. The hydrostatic
pressure scales as ${\rm P}_{\rm gas, hydro}$=7$\times$10$^{-10}$
$\Sigma_{\rm gas,500} \Sigma_{\rm total,1500}$ dyne cm$^{-2}$, where
$\Sigma_{\rm gas,500}$ is the gas mass surface density in units of 500
M$_{\sun}$ pc$^{-2}$ and $\Sigma_{\rm total,1500}$ is the stellar mass
surface density in units of 1500 M$_{\sun}$ pc$^{-2}$. For this estimate
of the hydrostatic pressure, we have assumed that the velocity dispersion
of the gas and stars are equal, which is consistent with our general
underlying hypothesis that the velocity dispersions of the warm ionized
and the cold molecular ISM phases are similar. However, this is not a
tight constraint as lowering the dispersion in the gas would lower the
turbulent pressure as well as the hydrostatic pressure for a fixed stellar
velocity dispersion. The estimated ranges of gas fraction and stellar
surface densities for z$\sim$2 galaxies \citep{daddi10,tacconi10} suggests
hydrostatic pressures of ${\rm P}_{\rm gas, hydro}$$\sim$10$^{-9}$
to 10$^{-10}$ dyne cm$^{-2}$. Thus it appears that the hydrostatic
pressure and the turbulent pressure in the molecular gas are similar if
the molecular gas has kinematics similar to that of the WIM.

\subsection{Surface densities and the Toomre Q parameter} 

We have argued that if the turbulence in the molecular gas were  similar
to that of the WIM, then the distant galaxies should lie near the line
of stability. We proposed a qualitative scenario where the kinematics of
the WIM may be roughly captured by the cold molecular gas and thus lead
to high turbulence. Furthermore, we suggested that these intensities can
be understood in terms of self-regulation \citep{silk97, silk01}. Such
self-regulation would reveal itself through mass surface densities in
the galaxies consistent with Q=1.

\citet{krumholz10} favor disk instabilities as the source of the
high velocity dispersion in distant galaxies with spatially-resolved
kinematics and line emission \citep[see also, e.g.][]{cacciato12}. Their
argument is that this is the simplest explanation, because it is based
on the Toomre instability parameter, Q=$\kappa\sigma$/$\pi$G$\Sigma$,
where $\kappa$ is the epicyclic frequency (which in our subsequent
analysis we have taken to be 2$^{1/2}$v/r, where v is the rotation
speed and r is the radius) and $\Sigma$ is the mass surface density
of stars or gas \citep[][]{wang94, romeo11}. They suggest that the
high velocity dispersions are a natural result of the need to keep the
star-formation intense, which implies that Q$\sim$1. The results given
above suggest perhaps that the star-formation itself may also lead to
Q$\sim$1 \citep[see also][]{ostriker11}. Combining the Schmidt-Kennicutt
relation with the Toomre instability criterion, Q=1, suggests that
$\sigma \propto \Sigma_{\rm SFR}^{0.7}$. This is similar enough to our
suggested relationship as to be indistinguishable \citep{wisnioski12,
swinbank12}. \citet{swinbank12} show that assuming a marginally
stable Toomre disk combined with relationships for the largest unstable
fluctuations in disks and a Jeans relation for the growth of unstable
modes, can lead to insights into why disks at high redshift may have high
dispersions, why clumps have high mass surface densities, why their
masses in disk galaxies decline with redshift and why their
mass distributions have exponential cutoffs.  However, such a solution
does not explain how the observed high dispersions are maintained. This
either requires a long dissipation time or a large source of mechanical
energy \citep{maclow99, ostriker11}. What maintains the instability such
that Q$\sim$1? We suggest it is the star-formation itself that maintains
the necessary dispersion in the gas phase, as we will now argue.

The Toomre parameter for the stellar population is
Q=$\kappa\sigma$/$\pi$G$\Sigma_{\rm stars}$, where $\Sigma_{\rm stars}$
is the mass surface density of stars. For the galaxies we have presented
here, the velocity shears are of-order 100-200 km s$^{-1}$ \citep{fs06,
fs09} and the radii are about a few to 10 kpc \citep{fs11}. The mass
surface density, $\Sigma_{\rm stars}$=M$_{\rm stars}$/2$\pi$(b/a)r$_e^2$,
where M$_{\rm stars}$ is the total stellar mass, b/a is the minor-to-major
axis ratio and r$_e$ is the effective radius. Using NICMOS and other
imaging data of 6 galaxies in our sample, we have made crude estimates of
stellar mass surface densities using the results from \citet{fs11}. We
find that the mass surface densities within one effective radius are
$\sim$100-1500 M$_{\sun}$ kpc$^{-2}$. Assuming Q=1, we can relate
the velocity dispersion to the mass surface density and it appears
that the estimated mass surface densities are consistent with Q=1
(Fig.~\ref{fig:toomreQ}).

Without considering the effects of beam smearing, this simple relation
is actually also roughly consistent with our data on the emission line
properties, as we can demonstrate rather simply: if we assume Q=1 for
the gas, and adopt the Schmidt-Kennicutt relation \citep{kennicutt98}
between star formation intensity and gas  surface density, we
find the relation, $\sigma$ = ($\pi$GQ/$\kappa$)($\Sigma_{\rm
SFR}$/2.5$\times$10$^{-4}$)$^{0.7}$. For a variety of rotation
speeds and disk radii, we find general consistency with our data
(Fig.~\ref{fig:toomreQSFR}). Of course these estimates are crude and
could be fortuitous but they do lend additional weight to our overall
arguments. We have not used the generalized (effective) Toomre criterion,
1/Q$_{\rm eff}$ = 1/Q$_{\rm gas}$ + 1/Q$_{\rm stars}$ \citep{wang94,jog96,
romeo11} because the crudeness of our analysis does not warrant this
additional complication. However, since this is a relatively small change
in the adopted Q, it will make little difference to our conclusions
which is that the galaxies generally lie near the line of stability.

\subsection{Toomre-stability of star-forming high-z
disks}\label{subsec:model}

Can we make a self-consistent model whereby intense star formation
is responsible for driving high turbulence and pushing the ISM in
distant galaxies towards the line of stability?  \citet{elmegreen10}
developed an analytical model to follow the evolution of gravitational
instabilities and star formation in disk galaxies that are built
from rapid gas accretion at high, constant rates. The energy of the
accretion is balanced by the dissipation rate and loss of energy due
to star formation. The accretion rate is defined by accreting half of a
fiducial mass scaling in one dynamical (orbital) time \citep[about 280
Myrs, ][]{elmegreen10}. Star formation only serves as an energy sink
in their formalism, to account for gas lost to the ISM because it is
bound in stars, but they do not include the energy return from young
stars into the ISM in a self-consistent way. In their model, turbulent
velocities drop to very low values well before stellar-to-gas mass
ratios reach observed values. To overcome this, they force turbulent
velocities, $\sigma$, to have values consistent with Q$=$1. They find
that ultimately the disk reaches an equilibrium where $\dot{\rm M}_{\rm
accretion}=\dot{\rm M}_{\rm star-formation}$.

\begin{figure}[ht!]
\includegraphics[width=9.0cm]{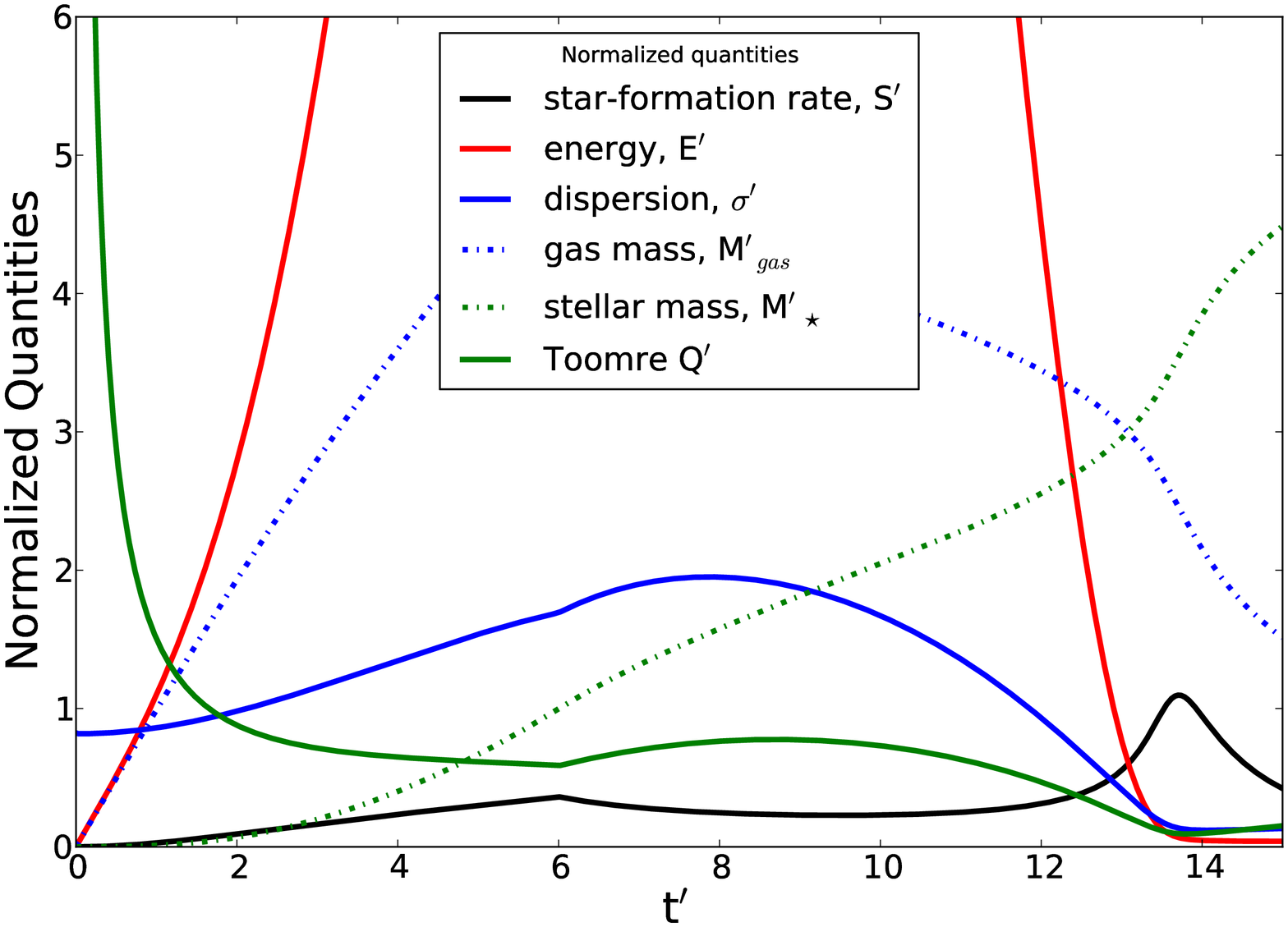}
\includegraphics[width=9.0cm]{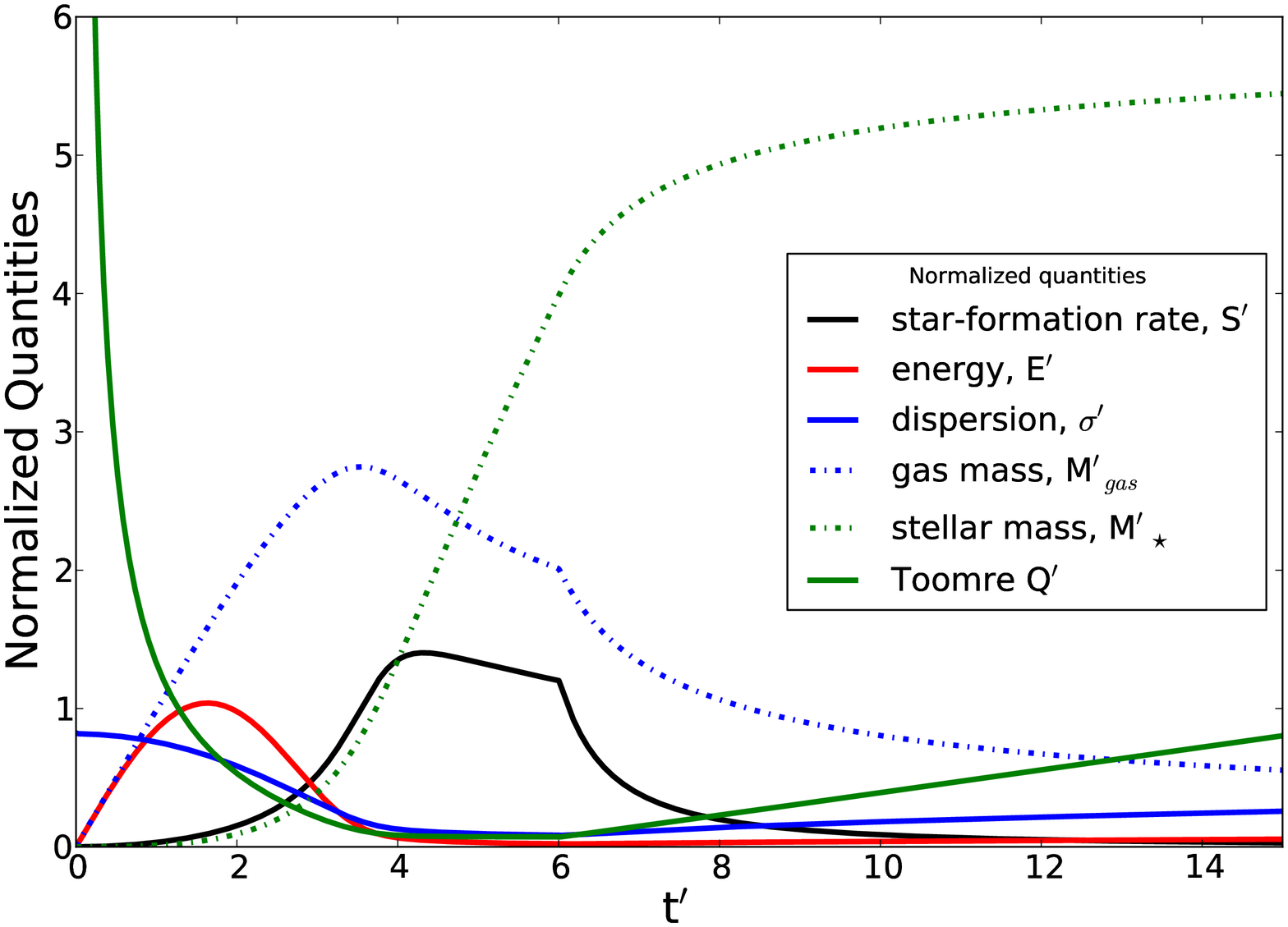}
\caption{Six normalized quantities (see legend) as a function of
time, where t/t$^{\prime}$ = 90 Myrs or about 1/3 of a dynamical
time of the potential. To limit the total mass so that it does not
become too large compared to the observations, the accretion ended
at t$^{\prime}$=6 (about 540 Myrs or two dynamical times). The top
panel is for a relatively high assumed efficiency of coupling of the
mechanical energy output of the stellar population to the ISM (12\%),
whereas  the bottom panel is for a relatively low coupling efficiency
(5\%). The scalings for these normalized quantities are provided in the
text \citep[see][]{elmegreen10}.}
\label{fig:models}
\end{figure}

We will now extend their model to include the effects of the energy
release of star formation. This is necessary if we want to investigate
how star formation is regulated in galaxies whose evolution is not
only driven by accretion and dissipation of turbulent energy, but
where star formation contributes significantly to the gas energetics,
as indicated by our observational finding that star formation produces
pressures similar to the mid-plane pressure of the disk. In our model,
accretion, dissipation and the energy released by star formation together
set Q. Including star-formation feedback in this way is one significant
difference to the model of \citet{elmegreen10}. The other is that we allow
for a cut-off in the gas accretion at some point in the evolution. Some
cut-off or severe reduction in the initial high rates of gas supply
must have occurred to produce  galaxies with  masses similar to those
generally observed at high redshift. This is not merely an academic
issue either. It is often argued that the star formation rate and the gas
accretion must be equal, and at the high star formation rates observed in
z$\sim$2 galaxies, the star formation time is a few 100 Myr to 1 Gyr. To
account for the energy from star formation, we have modified equation
(11) from \citet{elmegreen10} as follows:

\begin{equation}
\nonumber
\frac{{\rm dE}^{\prime}}{{\rm dt}^{\prime}} = 1 - {\rm M}^{\prime}_{{\rm disk}}({\rm E}^{\prime}{\rm M}^{\prime}_{{\rm gas}})^{1/2} - {\rm S}^{\prime}{\rm E}^{\prime}/{\rm M}^{\prime}_{{\rm gas}}\\ + \epsilon_{\rm SF} {\rm E}_{\rm SF} {\rm S}^{\prime} {{\rm M}^{\prime}_{{\rm gas}}}^{3/2} {{\rm E}^{\prime}}^{1/2}
\end{equation}

\noindent
where ${\rm E}^{\prime}$, ${\rm M}^{\prime}_{{\rm disk}}$, ${\rm
M}^{\prime}_{{\rm gas}}$, and ${\rm S}^{\prime}$ are the scaled
energy, disk mass, gas mass and star-formation rate, respectively
\citep[see][for details]{elmegreen10}. The first 3 terms are straight
from \citet{elmegreen10}. The first term is constant, because for
simplicity we assume a constant potential, and it is unity, because the
energy is normalized relative to the total accretion energy, which is
extracted from the potential. The second term quantifies the turbulent
dissipation. The dissipation rate is E$\sigma$/H, where E is the total
energy, $\sigma$ is the turbulent velocity and H is the scale height of
the disk, which  is determined by a competition between the turbulent
energy and the gravitational force due to the total mass surface density,
i.e., H=$\sigma^2$/$\pi$G$\Sigma_{disk}$. The third term is the energy
that is lost to the ISM when stars form. Adding a time dependence to
the potential is quite straight-forward and we have conducted both
a time independent and time dependent analysis. To these first three
terms from \citet{elmegreen10} we have added another term to incorporate
the energy injection rate from star formation, $\epsilon_{\rm SF} {\rm
E}_{\rm SF} {\rm S}^{\prime} {{\rm M}^{\prime}_{{\rm gas}}}^{3/2} {{\rm
E}^{\prime}}^{1/2}$, where $\epsilon_{\rm SF}$ is the coupling efficiency
of the mechanical energy to the interstellar medium of the galaxy and
${\rm E}_{\rm SF}$ is the rate at which energy is ejected per unit of
star formation. The whole equation is normalized such that the accretion
energy is one. In addition, other than the parameters already mentioned,
other key parameters in determining the evolution of this simple model
are the star formation efficiency, the gas accretion rate, the fraction
of the accretion energy that couples to the kinematics of the ISM, the
fraction of mechanical energy produced by young massive stars that cycles
back into the ISM and the dissipation timescale. For these parameters,
we adopted the  values from \citet{elmegreen10}, although we allowed
for a star formation efficiency of 3\% in the models instead of 1\%
they adopted. The fraction of energy produced by massive stars that
dynamically coupled to the ISM was a free parameter, which we would
like to constrain, and the dissipation timescale was determined by the
parameters of the model, as just discussed.

The scalings within this model are quite important. Using the estimates
of the mechanical energy per unit star formation \citep{leitherer99} and
a rotation speed of 220 km s$^{-1}$ (which sets the potential energy of
the galaxy from which the accretion energy is extracted), we estimate
that the mechanical energy injected into the ISM by stars is several
100 times greater than that generated by accretion. Also the time scale
for star formation is much shorter than that for accretion.  Thus star
formation can play an important role in injecting energy into the gas
and perhaps even pushing disks towards marginal stability against star
formation (Q$\sim$1). The scalings for the original model \citep[taken
from][]{elmegreen10} are t/t$^{\prime}$ = 90 Myrs (or about one-third
of an orbital time), S/S$^{\prime}$ = 160 M$_{\sun}$ yr$^{-1}$, M$_{\rm
gas}$/M$_{\rm gas}^{\prime}$ = 1.4 $\times$10$^{10}$ M$_{\sun}$ and
$\sigma$/$\sigma^{\prime}$=70 km s$^{-1}$. The scaled Toomre parameter,
Q$^{\prime}$ is related to the actual Toomre parameters by a factor of
$\approx$1/2 to 1, depending on the efficiency at which accretion energy
and the mechanical output from stars is converted into turbulence.

The results of this model are shown in Fig.~\ref{fig:models}. When
the coupling efficiency of the mechanical energy from the young stellar
population is high, around 5-15\%, Q$\sim$1 through most of the evolution
and because of this, the star-formation rate is roughly constant. The
energy injection into the ISM from the accretion and star formation is
high and therefore, dispersions are also high throughout much of the
evolution. With the adopted scalings, the dispersions are about 50-150
km s$^{-1}$, and star-formation rates are about 50-200 M$_{\sun}$
yr$^{-1}$. All of these values are within the range observed in our
sample. Star formation increases the Toomre parameter to Q$\sim$1
throughout the evolution and stabilizes disk star formation. At higher
coupling efficiencies, the gas dispersion can be even higher but the
star formation is suppressed. \citet{elmegreen10} set Q=1 when Q in
their model dropped below one. However, since we allow the energy
injection from star formation to evolve with star formation, we find a
different time evolution. Unlike in the model of \citet{elmegreen10},
the star-formation rate does not reach the accretion rate until well
after the accretion has stopped and the gas fraction has dropped below
50\% of the total (gas plus stellar) mass.

When the coupling  with the mechanical energy output of the
stellar population is weak, the disk is no longer stabilized near
Q$\sim$1 and Q becomes very small \citep[similar to the results
of][]{elmegreen10}. Consequently, star formation increases rapidly
and exceeds the accretion rate. This high star-formation rate occurs
again when the gas fraction drops below about 50\%. In this model,
star-formation rates are $\sim$200 M$_{\sun}$ yr$^{-1}$. This is high
but reasonable compared to the star-formation rates in our galaxies. The
values in Table~\ref{table:properties} are generally lower than this, but
they have not been corrected for extinction, which is likely to increase
them by factors of a few \citep{L09}. We note that during the early,
gas-rich phase of evolution in these models, the Jeans mass is relatively
high which implies that the disks would be very clumpy, especially if
the mechanical energy from stars couples very efficiently to the gas
\citep[because such galaxies stay relatively gas rich; ][]{elmegreen04,
elmegreen09a,elmegreen09,bournaud07,bournaud09, cervino10}.

The basic point illustrated by this simple model is that
star formation can play a key role in regulating large-scale
instabilities within the disks of distant galaxies. Previous studies
have come to the same conclusion in a somewhat different context
\citep[e.g.,][]{silk97,silk01,silk03, silk09}. What we have added to
the analysis is the use of the star formation to drive the line widths
(which are a combination of turbulent and bulk motions). In addition,
given a sufficiently high coupling efficiency, young stars can control
the subsequent star formation in distant galaxies. There does not need to
be correspondence between the instantaneous cosmological gas accretion
rate and the star-formation rate. In fact, quite the opposite. Smooth
constant accretion is probably unrealistic \citep[e.g.][]{danovich12}
and what we have shown here is that star formation, even if initiated
by the accretion event, does not have to be related closely in time with
accretion. The amount of energy available from the supernovae and stellar
winds is much larger in comparison and very early in the evolution begins
to control the star-formation history.

\subsection{Is the coupling unrealistically efficient?}\label{coupling}

There are  of course the important questions of whether or not
coupling efficiencies as high as those that are needed to explain
the large line widths are realistic and if the energy output from
the young stellar population through SNe and stellar winds can drive
the necessary turbulence. Some modeling studies suggest  this is not
the case \citep{d09, krumholz10}, while other suggest that it may be
\citep{joung09}.

Observationally, there are few constraints on what the value of the
coupling factor might be. \citet{krumholz10} criticized the analysis of
\citet{L09}, arguing that the required coupling efficiency between the
mechanical energy from star-formation and the ambient ISM is too high. Our
analysis in the previous section suggested that coupling efficiencies
of 5-15\% are necessary to explain the approximate characteristics of
distant galaxies, especially the relatively high velocity dispersions. But
even these estimates are very uncertain in that we do not really know
the dissipation time scales and that changes in the dissipation rate
will modify  the coupling factor. They must be short, approximately
one turnover time scale of turbulence (a few to tens of Myrs), as
argued in \citet{L09} and as suggested by the model discussed in the
previous section and the simulations and physical arguments presented
in \citep[e.g.][]{maclow99, elmegreen04, joung09}. The model suggests
that the dissipation time scale decreases with time and increases with
coupling efficiency. Since the dissipation rate is proportional to
the ratio of the velocity dispersion and the scale height of the disk,
the disk must be thick throughout much of its evolution. This agrees
with our finding that the gas has a relatively low ionization (and weak
radiation pressure) so it must be widely distributed in the ISM.

While we call this the ``coupling factor'', there are various  factors
that contribute to this coupling. To unwind these, we will use the well
studied nucleus of M82 as the analog for our distant galaxies, given that
the star formation intensity of M82 on a several 100 pc scale is similar
to that of the galaxies presented here. First, the mass ejection from
the stellar population must be efficiently thermalized. In a study of the
nuclear X-ray emission properties, which probes gas up to temperatures of
about 10$^{7.8}$K and directly probes the ``piston'' driving outflows,
\citet{strickland09} find that on scales of kpc, the thermalization
efficiency is 0.3-1.0. The entrainment factor -- the amount of ambient
ISM that is swept-up, accelerated, and mixed into the hot high pressure
plasma -- is also important. Again, the modeling of \citet{strickland09}
suggests relatively low entrainment rates of 1-4 times the mass loss of
the stellar population in the region surrounding the nucleus of M82. In
addition, in the optical emission line gas of M82, large line widths and
high pressures are observed \citep{west09}. This suggests that a fraction
of the energy from the intense star formation is dissipated in the WIM.

There are several ways in which the coupling between the thermalized
mechanical energy from young stars may be enhanced. Star formation which
is distributed increases the coupling factor between the mechanical
energy and the cold/warm ambient medium \citep{fragile04, cooper09}.
This enhancement results from the fact that with distributed regions
of star formation, say in a clumpy disk, overlapping bubbles of hot
plasma begin to interact and cause the formation of dense rims of gas
in this interaction zone \citep{T-T03, T-T06}. This also has the effect
of reducing the opening angles through which the hot plasma can flow
\citep{T-T03} which would thereby increase the central pressures in the
star forming regions, further enhancing its influence on the surrounding
ISM. Our photoionization and simple dynamical modeling implies that
our galaxies have a geometrically ``thick'' ($\sim$1 kpc) warm ionized
medium of moderate densities and low ionization parameters. Such a thick
medium would enhance the interaction between the mechanical luminosity
of the young stars and the ambient ISM. The strong interaction zones,
the reduced opening angle of the outflows, and a thick warm ionized
medium would have the effect of smothering the outflow and increasing
the efficiency at which the mechanical energy couples to the ambient
ISM, and also likely increases the cooling of the hot plasma generated
by the collective action of SNe and stellar winds in each star forming
clump or region \citep{wunsch08,hopkins12}. There is evidence that the
optical emission line gas in distant galaxies may be geometrically thick
\citep[e.g.]{epinat12}. In addition, these interacting and overlapping
regions of strong mechanical energy input may also drive gas to collapse
and  form molecular clouds which may be sites of subsequent star formation
\citep{T-T05, hennebelle08}, thus aiding in completing one part of the
gas cycle we are advocating.

Furthermore, we have up to now only considered the influence of the
mechanical energy output of the massive stars but not their radiation
field (except in the WIM itself). Depending on the star formation history
and duration, a significant fraction of up to many times the momentum
imparted by the mechanical output of a starburst could be imparted on
dusty gas by radiation pressure \citep{leitherer99,murray05}. Moreover,
if the optical depth to IR photons is high, the radiation pressure may
dominate the total momentum output \citep{hopkins11}. However, it remains
difficult to judge the importance of radiation pressure in  the overall
pressure \citep[cf.][]{murray05, silk10, ostriker11, hopkins11}. For our
discussion, the important point is that radiation pressure is more likely
to be effective in imparting momentum to the dense, dusty molecular gas
whereas the mechanical energy output will tend to flow around, engulf,
and destroy the molecular clouds. Thus radiation pressure may be very
effective in driving the large scale kinematics of the denser gas and
thus regulate star formation globally.

The coupling efficiency needed to drive the turbulence is dependent on
the energy dissipation timescale of the gas. In the range of coupling
efficiencies that lead to reasonable results compared to observations,
the energy dissipation timescale is in the range of a few to tens of
Myrs or in other words, shorter than a dynamical time or the duration
of the intense star formation, and approximately as long as one
turnover timescale of turbulence for a 1 kpc thick \Ha\ emitting layer
\citep[see][]{L09}. Thus in such a model, the cycle proposed is predicted
to be long-lived through repeated cycles of a turbulent cascade driven
by the energy input of massive stars. Since the rotational lines of
H$_2$ are one of the dominant coolants of warm molecular gas, and trace
the dissipation of supersonic turbulence \citep[e.g.][]{guillard10},
observations of distant galaxies with MIRI on the JWST in the infrared
will be crucial in constraining the coupling efficiency and testing the
scenario we have proposed.

\subsection{Summary}

Through our analysis of near-infrared integral field spectroscopic
observations we found that the optical emission line gas has a high
thermal pressure compared to nearby galaxies, and ionization parameters
that range from those of giant \HII\ regions to those of diffuse low
density gas away from star formation regions, similar to what is observed
in nearby normal and starburst galaxies \citep[e.g.][]{lehnert96, wang98,
LT11b, kewley02, pellegrini11}. The broad optical emission lines observed
($\sigma_{\Ha}$$\la$50-250 km s$^{-1}$) are not mainly due to beam
smearing but show that lines are intrinsically broader in regions of
more intense star formation. This suggests that the warm ionized gas
has a high level of turbulence and bulk motions, and participates in
outflows \citep{LT11b, genzel11} that are largely driven by the intense
star formation. The high level of turbulence and the influence of intense
star formation is consistent with the clumpy nature of distant galaxies
\citep[e.g.][]{elmegreen04,elmegreen09a,elmegreen09}.

The total pressure of the warm ionized medium appears to be dominated
by the thermal pressure as its average volume-weighted density is very
low. The relatively low ionization parameters of the emission line gas
suggest the thermal pressure in the WIM is due to the mechanical energy
output of massive stars and not radiation pressure of the ionizing field.

The evidence we have presented favors the high turbulence being driven
by mechanical output of young stars, which is the energy source that
regulates the gas to be near the line of disk instability (Q$\sim$1). The
mechanical energy from the intense star-formation is likely efficiently
thermalized \citep{strickland09}, creating large regions of high
turbulence. The intense radiation field from young stars may also impart
momentum to the gas, depending on the gas column densities and opacity
to IR photons. The importance of the radiation pressure is that it may
couple to the dense dusty gas more effectively than the mechanical energy
output, and thus regulate star formation globally.

Both the mechanical energy (through the thermalized hot plasma generated
by stellar winds and supernovae) and the radiation field of young stars
generate turbulence, bulk motions and high turbulent pressures in the gas
which are similar to the thermal pressure of the WIM and the hydrostatic
pressure. However, for the turbulent pressure to be similar to the
hydrostatic pressure, the high velocity dispersions observed in the
WIM have to be (at least partially) transferred to the dense molecular
gas, which likely has a much larger mass and average volume weighted
density than the recombining gas. Such transfer may happen because at
high pressures the recombining gas is unstable and will quickly cool
and collapse to a cold neutral phase \citep{wolfire95}, and become
molecular if dusty \citep{guillard09}. Since this collapse and cooling
time scale is much shorter than the dynamical time scale, the kinematics
of the WIM and the CNM should be similar \citep{guillard12}. This leads
to the formation of gravitationally bound clouds which have dissipated
their gravitational energy as turbulence in the molecular gas. Some of
these clouds will form stars, and some may be destroyed by the intense
energy output of massive stars. The ISM will quickly reach a dynamical
equilibrium, due to the relatively short timescales for gas cooling,
and the formation of molecular gas and gravitationally bound clouds. The
short timescale compared to those of star formation  indicates there is
sufficient time for repeated cycles, allowing for the build-up of the
stellar population in our distant galaxies.

To explain the observed star formation intensity -- line width
distribution, we made a simple model for the energy flow in a galaxy
undergoing both intense gas accretion and star formation. This model
suggests that only a few to perhaps 15\% of the injected energy from
stars is  required to drive the high level of turbulence we observe. This
relatively  small fraction of the total energy injection is reasonable.

If the high dispersion observed in the optical emission line gas is
due to turbulence and if the denser phases of the ISM share such high
turbulence, then the gas and stellar mass surface densities in z$\sim$2
galaxies are consistent with Q$\sim$1 and star formation is a plausible
hypothesis for driving the gas towards the line of stability. If so,
this would be an avenue for self-regulated star formation in distant
galaxies. More observations, especially of the warm molecular gas in
distant intensely star forming galaxies  are needed to gauge whether or
not this picture is correct.

\begin{acknowledgements} 

The work of LLT, MDL and PDM is supported by a grant from the Agence
Nationale de la Recherche (ANR) in France. MDL and NPHN wish to thank
the CNRS for their continuing support.

\end{acknowledgements}

\bibstyle{aa}
\bibliographystyle{aa}

\bibliography{L13v1}

\longtab{1}{
\begin{center}
\begin{longtable}{lclcccccc}
\caption{\label{table:properties} Properties of the High-z Galaxies}\\
\hline\hline
Object & Int & Line & z & $\sigma$ & Flux & SFR & SB limit & r$_{\rm iso}$ \\
(1) & (2) & (3) & (4) & (5) & (6) & (7) & (8) & (9) \\
\hline
\endfirsthead
\caption{continued.}\\
\hline\hline
Object & Int & Line & z & $\sigma$ & Flux & SFR & SB limit & r$_{\rm iso}$ \\
(1) & (2) & (3) & (4) & (5) & (6) & (7) & (8) & (9) \\
\hline
\endhead
\hline

\endfoot
BzK-15504$^b$ & 20400 & H$\alpha$ & 2.3816$\pm$0.0014 & 110$\pm$ 5 & 5.96$\pm$ 0.27 & 209$\pm$ 10 & 2.1 & 5.8 \\
 & & [N{\sc ii}]$\lambda$6583 & 2.3818$\pm$0.0016 & 134$\pm$ 12 & 2.44$\pm$ 0.20 & \\

BzK-15504 & 20400 & H$\alpha$ & 2.3822$\pm$0.0020 & 164$\pm$ 7 & 7.10$\pm$ 0.27 & 249$\pm$ 10 & 5.8 & 9.3 \\
 & & [N{\sc ii}]$\lambda$6583 & 2.3819$\pm$0.0016 & 133$\pm$ 14 & 1.96$\pm$ 0.21 & \\

BzK-6004 & 36000 & H$\alpha$ & 2.3865$\pm$0.0013 & 105$\pm$ 5 & 4.31$\pm$ 0.19 & 152$\pm$ 7 & 4.2 & 8.7 \\
 & & [N{\sc ii}]$\lambda$6583 & 2.3866$\pm$0.0010 & 67$\pm$ 6 & 1.77$\pm$ 0.15 & \\

BzK-6397 & 9600 & H$\alpha$ & 1.5133$\pm$0.0012 & 121$\pm$ 6 & 2.79$\pm$ 0.12 & 32$\pm$ 2 & 2.3 & 10.5 \\
 & & [N{\sc ii}]$\lambda$6583 & 1.5136$\pm$0.0016 & 178$\pm$ 17 & 1.14$\pm$ 0.10 & \\

CDFS-GK1084 & 7200 & H$\alpha$ & 1.5521$\pm$0.0009 & 87$\pm$ 8 & 0.38$\pm$ 0.03 & 4$\pm$ 1 & 2.9 & 4.4 \\
 & & [N{\sc ii}]$\lambda$6583 & 1.5527$\pm$0.0009 & 88$\pm$ 20 & 0.14$\pm$ 0.03 & \\

CDFS-GK2252 & 18000 & H$\alpha$ & 2.4080$\pm$0.0014 & 111$\pm$ 9 & 0.90$\pm$ 0.06 & 32$\pm$ 3 & 7.8 & 4.7 \\
 & & [NII]$\lambda$6583 & 2.4073$\pm$0.0013 & 101$\pm$ 28 & 0.21$\pm$ 0.06 & \\

CDFS-GK2363 & 17400 & H$\alpha$ & 2.4512$\pm$0.0014 & 105$\pm$ 8 & 1.13$\pm$ 0.09 & 42$\pm$ 4 & 8.1 & 4.4 \\ 
 & & [NII]$\lambda$6583 & 2.4508$\pm$0.0009 & 60$\pm$ 15 & 0.25$\pm$ 0.06 & \\

CDFS-GK2471 & 16800 & H$\alpha$ & 2.4319$\pm$0.0019 & 159$\pm$ 8 & 2.20$\pm$ 0.11 & 81$\pm$ 5 & 8.3 & 5.3 \\ 
 & & [NII]$\lambda$6583 & 2.4350$\pm$0.0005 & \ldots & $<$0.34$^c$ & \\

DEEP2-32002481 & 6600 & H$\alpha$ & 1.3880$\pm$0.0014 & 156$\pm$ 19 & 2.77$\pm$ 0.32 & 25$\pm$ 3 & 2.5 & 9.8 \\ 
 & & [NII]$\lambda$6583 & 1.3897$\pm$0.0017 & 209$\pm$ 65 & 1.20$\pm$ 0.37 & \\

DEEP2-32007614 & 5400 & H$\alpha$ & 1.3716$\pm$0.0010 & 110$\pm$ 8 & 1.12$\pm$ 0.08 & 10$\pm$ 1 & 6.9 & 3.9 \\ 
 & & [NII]$\lambda$6583 & 1.3718$\pm$0.0006 & 42$\pm$ 6 & 0.46$\pm$ 0.06 & \\

DEEP2-32100778 & 6000 & H$\alpha$ & 1.3918$\pm$0.0006 & 41$\pm$ 4 & 3.54$\pm$ 0.27 & 33$\pm$ 3 & 3.9 & 8.8 \\ 
 & & [NII]$\lambda$6583 & 1.3919$\pm$0.0007 & 59$\pm$ 14 & 1.36$\pm$ 0.31 & \\

DEEP2-32013051 & 5400 & H$\alpha$ & 1.3950$\pm$0.0011 & 118$\pm$ 9 & 2.45$\pm$ 0.18 & 23$\pm$ 2 & 2.8 & 7.7 \\ 
 & & [NII]$\lambda$6583 & 1.3951$\pm$0.0011 & 124$\pm$ 16 & 1.57$\pm$ 0.19 & \\

DEEP2-32015443 & 6600 & H$\alpha$ & 1.3822$\pm$0.0009 & 85$\pm$ 13 & 1.48$\pm$ 0.22 & 13$\pm$ 3 & 6.0 & 6.7 \\ 
 & & [NII]$\lambda$6583 & 1.3824$\pm$0.0009 & 83$\pm$ 28 & 0.54$\pm$ 0.18 & \\

DEEP2-32015501 & 10800 & H$\alpha$ & 1.3919$\pm$0.0007 & 58$\pm$ 5 & 2.47$\pm$ 0.18 & 23$\pm$ 2 & 3.8 & 9.5 \\ 
 & & [NII]$\lambda$6583 & 1.3921$\pm$0.0006 & 32$\pm$ 12 & 0.38$\pm$ 0.14 & \\

DEEP2-32021317 & 5400 & H$\alpha$ & 1.3815$\pm$0.0017 & 198$\pm$ 20 & 2.92$\pm$ 0.28 & 26$\pm$ 3 & 4.3 & 8.3 \\ 
 & & [NII]$\lambda$6583 & 1.3822$\pm$0.0028 & 348$\pm$ 39 & 3.34$\pm$ 0.37 & \\

DEEP2-32021394 & 5400 & H$\alpha$ & 1.3735$\pm$0.0009 & 92$\pm$ 9 & 2.78$\pm$ 0.25 & 25$\pm$ 3 & 4.0 & 9.5 \\ 
 & & [NII]$\lambda$6583 & 1.3753$\pm$0.0013 & 156$\pm$ 34 & 1.29$\pm$ 0.28 & \\

DEEP2-32029850 & 6600 & H$\alpha$ & 1.3952$\pm$0.0009 & 81$\pm$ 22 & 0.13$\pm$ 0.03 & 1$\pm$ 1 & 3.6 & 2.5 \\ 
 & & [NII]$\lambda$6583 & 1.3960$\pm$0.0010 & 108$\pm$ 36 & 0.09$\pm$ 0.03 & \\

DEEP2-32037003 & 5400 & H$\alpha$ & 1.3985$\pm$0.0014 & 157$\pm$ 7 & 5.50$\pm$ 0.24 & 52$\pm$ 3 & 3.6 & 9.6 \\ 
 & & [NII]$\lambda$6583 & 1.3987$\pm$0.0010 & 103$\pm$ 15 & 1.40$\pm$ 0.19 & \\

ECDFS-10525 & 5400 & H$\alpha$ & 2.0261$\pm$0.0029 & 278$\pm$ 27 & 6.90$\pm$ 0.67 & 163$\pm$ 16 & 11.5 & 7.8 \\ 
 & & [NII]$\lambda$6583 & 2.0271$\pm$0.0014 & 118$\pm$ 20 & 2.59$\pm$ 0.43 & \\

ECDFS-2896 & 3600 & H$\alpha$ & 2.3060$\pm$0.0017 & 137$\pm$ 17 & 1.55$\pm$ 0.19 & 50$\pm$ 7 & 16.0 & 3.5 \\ 
 & & [NII]$\lambda$6583 & 2.3050$\pm$0.0019 & 155$\pm$ 29 & 1.14$\pm$ 0.21 & \\

F257 & 21600 & H$\alpha$ & 2.0264$\pm$0.0014 & 122$\pm$ 11 & 1.92$\pm$ 0.16 & 45$\pm$ 4 & 2.8 & 9.3 \\ 
 & & [NII]$\lambda$6583 & 2.0255$\pm$0.0011 & 89$\pm$ 29 & 0.38$\pm$ 0.12 & \\ 

K20-ID4 & 9900 & H$\alpha$ & 1.5934$\pm$0.0008 & 62$\pm$ 4 & 2.85$\pm$ 0.14 & 37$\pm$ 2 & 1.7 & 8.3 \\ 
 & & [NII]$\lambda$6583 & 1.5937$\pm$0.0008 & 64$\pm$ 7 & 0.79$\pm$ 0.08 & \\

K20-ID5 & 12000 & H$\alpha$ & 2.2240$\pm$0.0022 & 195$\pm$ 10 & 2.97$\pm$ 0.14 & 88$\pm$ 5 & 5.6 & 6.1 \\ 
 & & [NII]$\lambda$6583 & 2.2240$\pm$0.0029 & 260$\pm$ 23 & 1.83$\pm$ 0.16 & \\

K20-ID7 & 25200 & H$\alpha$ & 2.2233$\pm$0.0017 & 148$\pm$ 7 & 4.38$\pm$ 0.18 & 130$\pm$ 6 & 2.9 & 9.4 \\ 
 & & [NII]$\lambda$6583 & 2.2243$\pm$0.0013 & 111$\pm$ 14 & 0.87$\pm$ 0.11 & \\

K20-ID8 & 15000 & H$\alpha$ & 2.2230$\pm$0.0011 & 89$\pm$ 5 & 2.20$\pm$ 0.11 & 65$\pm$ 4 & 3.7 & 7.5 \\ 
 & & [NII]$\lambda$6583 & 2.2241$\pm$0.0015 & 123$\pm$ 17 & 0.74$\pm$ 0.10 & \\

Q1623-BX376 & 17400 & H$\alpha$ & 2.4080$\pm$0.0011 & 83$\pm$ 6 & 0.36$\pm$ 0.02 & 12$\pm$ 1 & 4.6 & 3.5 \\
 & & [NII]$\lambda$6583 & 2.4085$\pm$0.0007 & 32$\pm$ 11 & 0.05$\pm$ 0.02 & \\

Q1623-BX447 & 13800 & H$\alpha$ & 2.1471$\pm$0.0014 & 118$\pm$ 11 & 1.07$\pm$ 0.10 & 29$\pm$ 3 & 4.2 & 5.8 \\
 & & [NII]$\lambda$6583 & 2.1498$\pm$0.0005 & \ldots & $<$0.35$^c$ & \\

Q1623-BX455 & 12000 & H$\alpha$ & 2.4065$\pm$0.0014 & 107$\pm$ 8 & 1.41$\pm$ 0.09 & 50$\pm$ 4 & 10.7 & 4.2 \\
 & & [NII]$\lambda$6583 & 2.4080$\pm$0.0015 & 124$\pm$ 34 & 0.35$\pm$ 0.09 & \\

Q1623-BX543 & 8400 & H$\alpha$ & 2.5204$\pm$0.0017 & 137$\pm$ 6 & 5.94$\pm$ 0.26 & 239$\pm$ 11 & 12.5 & 6.1 \\
 & & [NII]$\lambda$6583 & 2.5160$\pm$0.0011 & 72$\pm$ 21 & 0.68$\pm$ 0.19 & \\

Q1623-BX599 & 5400 & H$\alpha$ & 2.3305$\pm$0.0019 & 160$\pm$ 7 & 5.99$\pm$ 0.24 & 199$\pm$ 9 & 8.2 & 6.2 \\
 & & [NII]$\lambda$6583 & 2.3302$\pm$0.0011 & 85$\pm$ 17 & 0.76$\pm$ 0.15 & \\

Q1623-BX663 & 33300 & H$\alpha$ & 2.4253$\pm$0.0026 & 219$\pm$ 12 & 1.96$\pm$ 0.11 & 72$\pm$ 5 & 4.3 & 6.2 \\
 & & [NII]$\lambda$6583 & 2.4265$\pm$0.0038 & 325$\pm$ 58 & 0.75$\pm$ 0.13 & \\

\clearpage

Q2343-BX389 & 14400 & H$\alpha$ & 2.1716$\pm$0.0020 & 180$\pm$ 7 & 4.59$\pm$ 0.17 & 128$\pm$ 5 & 3.4 & 8.5 \\
 & & [NII]$\lambda$6583 & 2.1727$\pm$0.0014 & 122$\pm$ 27 & 0.69$\pm$ 0.15 & \\

Q2343-BX502 & 9600 & H$\alpha$ & 2.1550$\pm$0.0008 & 52$\pm$ 3 & 1.82$\pm$ 0.09 & 50$\pm$ 3 & 2.8 & 7.7 \\
 & & [NII]$\lambda$6583 & 2.1571$\pm$0.0010 & \ldots & 0.23$^d$ & \\

Q2343-BX528 & 34200 & H$\alpha$ & 2.2684$\pm$0.0015 & 128$\pm$ 6 & 2.26$\pm$ 0.10 & 70$\pm$ 4 & 4.2 & 7.5 \\
 & & [NII]$\lambda$6583 & 2.2691$\pm$0.0010 & 72$\pm$ 10 & 0.41$\pm$ 0.05 & \\

Q2343-BX610 & 21600 & H$\alpha$ & 2.2098$\pm$0.0018 & 163$\pm$ 7 & 7.91$\pm$ 0.32 & 231$\pm$ 10 & 3.7 & 9.3 \\
 & & [NII]$\lambda$6583 & 2.2105$\pm$0.0013 & 106$\pm$ 6 & 2.49$\pm$ 0.13 & \\

Q2346-BX482 & 31200 & H$\alpha$ & 2.2562$\pm$0.0015 & 131$\pm$ 6 & 4.09$\pm$ 0.17 & 126$\pm$ 6 & 3.4 & 9.0 \\
 & & [NII]$\lambda$6583 & 2.2566$\pm$0.0010 & 70$\pm$ 10 & 0.48$\pm$ 0.06 & \\

SA12-5241 & 7200 & H$\alpha$ & 1.3617$\pm$0.0007 & 53$\pm$ 3 & 1.46$\pm$ 0.07 & 13$\pm$ 1 & 4.3 & 6.0 \\
 & & [NII]$\lambda$6583 & 1.3619$\pm$0.0006 & 37$\pm$ 4 & 0.62$\pm$ 0.06 & \\

SA12-6192 & 10800 & H$\alpha$ & 1.5038$\pm$0.0008 & 62$\pm$ 5 & 0.69$\pm$ 0.05 & 7$\pm$ 1 & 1.8 & 5.7 \\
 & & [NII]$\lambda$6583 & 1.5038$\pm$0.0009 & 84$\pm$ 12 & 0.40$\pm$ 0.06 & \\

SA12-6339 & 19200 & H$\alpha$ & 2.2963$\pm$0.0013 & 102$\pm$ 5 & 3.33$\pm$ 0.15 & 107$\pm$ 5 & 4.3 & 6.0 \\
 & & [NII]$\lambda$6583 & 2.2950$\pm$0.0016 & 129$\pm$ 26 & 0.61$\pm$ 0.12 & \\

SA12-8768 & 10800 & H$\alpha$ & 2.1875$\pm$0.0010 & 73$\pm$ 4 & 2.31$\pm$ 0.12 & 66$\pm$ 4 & 5.5 & 6.6 \\
 & & [NII]$\lambda$6583 & 2.1875$\pm$0.0011 & 85$\pm$ 21 & 0.44$\pm$ 0.10 & \\

SA15-5365 & 9000 & H$\alpha$ & 1.5340$\pm$0.0011 & 117$\pm$ 12 & 1.10$\pm$ 0.10 & 13$\pm$ 2 & 2.1 & 6.8 \\
 & & [NII]$\lambda$6583 & 1.5323$\pm$0.0010 & 100$\pm$ 40 & 0.19$\pm$ 0.07 & \\

SINS-4751 & 3600 & H$\alpha$ & 2.2651$\pm$0.0011 & 78$\pm$ 4 & 3.12$\pm$ 0.16 & 97$\pm$ 5 & 6.2 & 5.8 \\
 & & [NII]$\lambda$6583 & 2.2651$\pm$0.0010 & 69$\pm$ 17 & 0.48$\pm$ 0.11 & \\

SSA-22a-MD041 & 21600 & H$\alpha$ & 2.1632$\pm$0.0014 & 122$\pm$ 6 & 3.54$\pm$ 0.16 & 98$\pm$ 5 & 2.9 & 8.3 \\
 & & [NII]$\lambda$6583 & 2.1627$\pm$0.0016 & 142$\pm$ 39 & 0.50$\pm$ 0.13 & \\
 
VVDS-020147106 & 7200 & H$\alpha$ & 1.5182$\pm$0.0008 & 76$\pm$ 4 & 5.37$\pm$ 0.27 & 62$\pm$ 4 & 4.1 & 7.6 \\ 
 & & [NII]$\lambda$6583 & 1.5183$\pm$0.0007 & 40$\pm$ 13 & 0.31$\pm$ 0.10 & \\

VVDS-020261328 & 3600 & H$\alpha$ & 1.5276$\pm$0.0008 & 71$\pm$ 4 & 1.00$\pm$ 0.06 & 11$\pm$ 1 & 7.9 & 4.3 \\
 & & [NII]$\lambda$6583 & 1.5279$\pm$0.0006 & 39$\pm$ 13 & 0.12$\pm$ 0.04 & \\

VVDS-220014252 & 7200 & H$\alpha$ & 1.3090$\pm$0.0011 & 130$\pm$ 6 & 7.84$\pm$ 0.35 & 63$\pm$ 3 & 5.6 & 8.5 \\ 
 & & [NII]$\lambda$6583 & 1.3090$\pm$0.0011 & 128$\pm$ 35 & 1.02$\pm$ 0.28 & \\

VVDS-220015726 & 10800 & H$\alpha$ & 1.2920$\pm$0.0012 & 140$\pm$ 7 & 6.63$\pm$ 0.30 & 51$\pm$ 3 & 5.2 & 7.7 \\
 & & [NII]$\lambda$6583 & 1.2919$\pm$0.0010 & 111$\pm$ 17 & 1.09$\pm$ 0.16 & \\

VVDS-220544103 & 7200 & H$\alpha$ & 1.3955$\pm$0.0010 & 103$\pm$ 5 & 7.77$\pm$ 0.35 & 73$\pm$ 4 & 5.6 & 8.7 \\ 
 & & [NII]$\lambda$6583 & 1.3960$\pm$0.0012 & 141$\pm$ 23 & 1.55$\pm$ 0.25 & \\

VVDS-220584167 & 7200 & H$\alpha$ & 1.4643$\pm$0.0012 & 125$\pm$ 6 & 9.37$\pm$ 0.42 & 99$\pm$ 5 & 2.6 & 11.5 \\ 
 & & [NII]$\lambda$6583 & 1.4645$\pm$0.0010 & 98$\pm$ 10 & 1.94$\pm$ 0.18 & \\

VVDS-220596913 & 10800 & H$\alpha$ & 1.2649$\pm$0.0010 & 107$\pm$ 5 & 4.11$\pm$ 0.18 & 30$\pm$ 2 & 3.3 & 8.5 \\ 
 & & [NII]$\lambda$6583 & 1.2654$\pm$0.0011 & 125$\pm$ 31 & 0.70$\pm$ 0.17 & \\

ZC-1101592 & 3600 & H$\alpha$ & 1.4034$\pm$0.0013 & 150$\pm$ 16 & 1.07$\pm$ 0.11 & 10$\pm$ 2 & 10.6 & 4.1 \\
 & & [NII]$\lambda$6583 & 1.4035$\pm$0.0006$^d$ & 7$\pm$ 2$^d$ & 0.24$\pm$ 0.07$^d$ & \\

ZC-782941$^b$ & 13800 & H$\alpha$ & 2.1813$\pm$0.0016 & 144$\pm$ 6 & 3.14$\pm$ 0.13 & 89$\pm$ 4 & 3.1 & 4.1 \\
 & & [NII]$\lambda$6583 & 2.1814$\pm$0.0010 & 77$\pm$ 10 & 0.74$\pm$ 0.09 & \\

ZC-782941$^a$ & 7200 & H$\alpha$ & 2.1811$\pm$0.0015 & 133$\pm$ 6 & 4.97$\pm$ 0.21 & 140$\pm$ 6 & 6.2 & 7.2 \\
 & & [NII]$\lambda$6583 & 2.1812$\pm$0.0010 & 76$\pm$ 19 & 0.71$\pm$ 0.17 & \\
\hline\\

\caption{Column (1) -- Object designation. $^a$ denotes a 250 mas pixel scale plus adoptive optics, and $^b$ a 100 mas pixel scale plus adaptive optics. No indication denotes that the data were taken without the benefit of AO, and at 250 mas pixel$^{-1}$;
Column (2) -- Integration time in seconds;
Column (3) -- Line identifications;
Column (4) -- Redshifts of the H$\alpha$ and [NII]$\lambda$6583 lines in the integrated spectrum. By integrated spectrum, we mean the sum of the fluxes from each pixel with an H$\alpha$ line signal-to-noise ratio $\ga$3 in a specific data cube (see column 8 for the 3$\sigma$ surface brightness detection limits). For each object, the sum for the [NII]$\lambda$6583 emission line was determined over the same aperture as for H$\alpha$;
Column (5) -- H$\alpha$ line velocity dispersion of the integrated spectrum of each galaxy, corrected for instrumental resolution and in units of km s$^{-1}$; 
Column (6) -- H$\alpha$ and [NII]$\lambda$6583 line fluxes of the integrated spectrum in units of 10$^{-16}$ erg s$^{-1}$ cm$^{-2}$. $^c$The [NII] line is not detected and the value given is the 3-$\sigma$ upper limit, for a line width equal to that of the H$\alpha$ line, and $^d$The estimated line characteristics are influenced significantly by a nearby night sky line;
Column (7) -- Star Formation Rate, in M$_{\sun}$ yr$^{-1}$, estimated from the total H$\alpha$ luminosity following \citep{kennicutt98}. These values have not been corrected for extinction, which is likely to increase them by a factor of a few (see text);
Column (8) -- H$\alpha$ surface brightness detection limit in units of 10$^{-19}$ erg s$^{-1}$ cm$^{-2}$, at a signal-to-noise level of S/N$\approx$3. These are average values over a 3 $\times$ 3 pixel area;
Column (9) -- Isophotal radius, r$_{\rm iso}$, defined as A$_{\rm iso}$ = $\pi$ r$_{\rm iso}^2$. The isophotal area, A$_{\rm iso}$, is defined as the projected area on the sky above a signal-to-noise level of 3 in the \Ha data. r$_{\rm iso}$ is in kpc.}
\end{longtable}
\end{center}}

\end{document}